# Charged-impurity free printing-based diffusion doping in molybdenum disulfide field-effect transistors

Inho Jeong[1†], Jiwoo Yang[1,2†], Juntae Jang[3†], Daeheum Cho[4], Deok-Hwang Kwon[5], Jae-Keun Kim[6], Takhee Lee[3]*, Kyungjune Cho[7,8]*, and Seungjun Chung[1]*

[1]School of Electrical Engineering, Korea University, Seoul, 02841, Republic of Korea

[2]Department of Electrical and Computer Engineering, Seoul National University, Seoul 08826, Republic of Korea

[3]Department of Physics and Astronomy, and Institute of Applied Physics, Seoul National University, Seoul 08826, Republic of Korea

[4]Departments of Chemistry, Kyungpook National University, Daegu 41566, Republic of Korea

[5]Energy Materials Research Center, Korea Institute of Science and Technology, Seoul 02792, Republic of Korea

[6]Max-Planck Institute of Microstructure Physics, Weinberg2, 06120 Halle, Saale, Germany.

[7]Soft Hybrid Materials Research Center, Korea Institute of Science and Technology, Seoul 02792, Republic of Korea

[8]Convergence Research Center for Solutions to Electromagnetic Interference in Future-mobility, Korea Institute of Science and Technology, Seoul 02792, Republic of Korea

Correspondence and requests for materials should be addressed to Takhee Lee (e-mail: tlee@snu.ac.kr), Kyungjune Cho (e-mail: kcho@kist.re.kr) and Seungjun Chung

(seungjun@korea.ac.kr).

[†]These authors contributed equally to this work.

## Abstract

In practical electronic applications, where doping is crucial to exploit large-area two-dimensional (2D) semiconductors, surface charge transfer doping (SCTD) has emerged as a promising strategy to tailor their electrical characteristics. However, impurity scattering caused by resultant ionized dopants, after donating or withdrawing carriers, hinders transport in 2D semiconductor layers, limiting the carrier mobility. Here, we propose a diffusion doping method for chemical vapor deposition (CVD)-grown molybdenum disulfide ($MoS_2$) that avoids interference from charged impurities. Selectively inkjet-printed dopants were introduced only on the contact region, allowing excessively donated electrons to diffuse to the channel layer due to the electron density difference. Therefore, diffusion-doped $MoS_2$ FETs do not have undesirable charged impurities on the channel, exhibiting over two-fold higher field-effect mobility compared with conventional direct-doped ones. Our study paves the way for a new doping strategy that simultaneously suppresses charged impurity scattering and facilitates the tailoring of the SCTD effect.

As demands for large-area, processable, and versatile solutions in modern electronic devices increase, two-dimensional (2D) transition metal dichalcogenides (TMDCs) have attracted significant interest due to their remarkable properties, including their ultra-thin 2D nature and excellent electronic and optoelectronic characteristics.[1–5]. These features position TMDCs as strong candidates for realising scalable electronic applications, such as field-effect transistors (FETs) and optoelectronic devices. Specifically, large-area monolayer molybdenum disulfide ($MoS_2$), an *n*-type TMDC semiconductor synthesised via chemical vapour deposition (CVD), holds significant potential for practical applications. However, to realise its full potential in real-world electronic devices beyond laboratory-scale implementation, efficient control of its intrinsic electrical properties is essential. Doping is a critical factor in modulating the electrical properties of semiconducting materials, including their carrier density and charge transport properties[4–15]. Therefore, numerous doping strategies have been developed to precisely control the electrical properties of TMDCs. Surface charge transfer doping (SCTD), which involves transferring charge between TMDCs and externally introduced dopants on their surfaces, has been widely investigated as an effective method for modulating the electrical properties of TMDCs. Owing to their intrinsic 2D nature, TMDCs offer a large surface-to-volume ratio leading much efficient surface charge transfer from adsorbed dopants without disrupting the atomic structure of 2D TMDCs [8,9,11,12,14,16–19].

One of the remaining challenges in SCTD is undesirable charged impurity scattering[7,15,20,21]. This issue arises when dopants in the SCTD process donate their charges to the adjacent 2D TMDCs, causing the dopants to acquire opposite charges, as illustrated in Fig. 1a. These charged impurities create a Coulomb potential barrier within the channel, hindering charge transport, reducing mobility, and posing a significant obstacle for practical high-speed 2D electronics[18,22,23].

Efforts have been made to mitigate the effects of charged impurity scattering. Because the Coulomb potential primarily depends on distance, increasing the physical thickness of 2D TMDCs has been regarded as a solution. By enlarging the distance between the charged impurities on the surface and charge carriers in the channel, this approach reduces the magnitude of the Coulomb potential. Therefore, several recent doping approaches have been proposed to minimise the effects of charged impurities. For example, Lee *et al.* suppressed charged impurity scattering by employing an additional band-modulated 2D semiconductor layer, which is physically separated from the target 2D semiconductor channel by an atomically thin *h*-BN layer[22]. Similarly, Jang *et al.* reported that inserting a 2–3 nm thick *h*-BN layer between the charge transfer dopants and 2D TMDC channel can reduce charged impurity scattering[23]. Although these remote doping strategies can effectively suppress charged impurity scattering, their implementation has been limited to mechanically exfoliated single flakes of van der Waals (vdWs) heterostructures produced through labourious processes. Achieving charged impurity-free doping in large-area monolayer TMDCs *via* remote doping remains a substantial challenge for practical applications owing to the difficulties of synthesising large-area *h*-BN and creating heterostructures with clean interfaces. In addition, the demand for offering further exquisitely tailored doping effects has increased for modern electronics as the operation window gets smaller, which cannot be achieved with complicated remote doping or conventional direct doping systems.

In this study, we introduce a new doping strategy called diffusion doping, specifically designed for large-area CVD-grown monolayer TMDC-based devices using single-step direct ink writing. This methodology effectively enables the donation of charge carriers to the underlying 2D TMDC channels while avoiding the introduction of charged impurities into the channel region. By utilising a large-area processable drop-on-demand inkjet printing technique, we achieved both spatial selectivity and facile tailoring of the doping effect in wafer-scale $MoS_2$ FETs using different concentrations of dopant inks with benzyl viologen (BV), a representative *n*-type molecular dopant for SCTD. The inkjet-printed BV dopants successfully donated a substantial number of electrons onto the only contact region of the semiconducting 2D TMDC layer. These excessive electrons then diffused to the channel region, resulting in desirable doping effects without the presence of physical dopants in the channel area (*i.e.* a charged impurity-free surface), as illustrated in Fig. 1c. Through a combination of various electrical and surface characterisations, along with theoretical calculations, we validated the effectiveness of our diffusion doping method. The density functional theory (DFT) calculations suggest that a large reduction in the Fermi level of $MoS_2$, along with a corresponding increase in the effective density of states (DOS) within the bandgap, can occur as the number of BV molecular dopant layers on $MoS_2$ increases (Fig. 1d). These results indicate that increasing the number of molecular dopant layers can enhance charge transfer efficiency, in turn allowing $MoS_2$ highly doped. We also confirmed the absence of dopants on the channel surface, even as the carrier concentration within the channel increases. Significantly, large-area inkjet printing facilitates selective dopant deposition onto CVD-grown monolayer $MoS_2$ films, focusing solely on the contact regions and providing a high degree of spatial freedom. This technique offers markedly superior precision to that achieved with conventional SCTD approaches, such as immersion or spin coating, without complicated patterning processes. We found that it successfully enables superior mobility in CVD-grown monolayer $MoS_2$ FETs, even in the low-temperature regime where charged impurity scattering is dominant.

The electrical properties of diffusion-doped CVD-grown $MoS_2$ FETs with deposited BV inks at different doping concentrations were evaluated (Fig. 2a–f). The transfer and output curves of pristine (black) and diffusion-doped $MoS_2$ FETs (red) are presented for concentrations ranging from 1 to 20 mg/mL to investigate the efficiency and controllability of the diffusion doping process. An increase in current levels in both on- and off-states is noted with higher doping concentrations. Concurrently, the turn-on voltage shifts in the negative gate voltage direction, indicating a higher carrier concentration within the channel, even when only BV dopants are introduced at the contact regions. Given that the total device resistance comprises two contact resistances and channel resistances, as depicted in Fig. 2g, the observed decrease in total device resistance necessarily involves a reduction in channel resistance. Since BV molecules are exclusively deposited at the contact region and the channel length is ~200 um, the reduced resistance can only be originated to the diffusion doping. Particularly, the $MoS_2$ FET doped with a 20 mg/mL ink concentration exhibits a substantial increase in off-current compared to its pristine counterpart (Fig. 2f), suggesting that BV deposition induces degenerate doping in the long channel, modulating off-state resistances from hundreds of GΩ to hundreds of KΩ. The observed changes in the electrical properties can be attributed to two key factors: enhanced carrier injection from the contacts into the $MoS_2$ layer and improved carrier diffusion from the contact region to the $MoS_2$ channel. Additionally, a comparable doping effect is noted in the case of a 5 mg dopant immersion, as detailed in Supplementary Information Section 4.

To validate the effectiveness of diffusion doping and ensure that it achieves the desired doping effect without BV molecules residing in the channel region, we optimised the printing conditions for stable and precise inkjet printing of BV molecules on CVD-grown $MoS_2$ films (Fig. S1). Subsequently, we confirmed the absence of BV molecules in the channel region through various analytical techniques, including optical microscopy, X-ray photoelectron spectroscopy (XPS), scanning electron microscopy (SEM), and atomic force microscopy (AFM). Further details are available in the Supplementary Information Section 1. Additionally, the integrity of the inkjet-printed BV molecules was maintained following the application of Ag ink, which was used for contact deposition on the 2 nm thick BV layer. This finding was corroborated by cross-sectional Energy-filtered transmission electron microscopy (Fig. S3 and S4).

Upon being deposited in the contact region, BV molecules are anticipated to transfer electrons to the underlying $MoS_2$ layer. Then, the high concentration of electrons at the contact regions facilitates electron diffusion until the potential energy barrier becomes substantial enough to inhibit further movement, as shown in Fig. 3a. This mechanism suggests the formation of an electron density gradient between the directly doped contact region and channel region owing to the transferred charges, a phenomenon also discernible in the DFT-calculated electron density profile as shown in Fig. 1c. Moreover, previous research revealed that excess charges from external sources can diffuse over several tens of micrometres in equilibrium states, a characteristic attributed to the low electron density of monolayer CVD-grown $MoS_2$[25]. Note that this behaviour differs from that of excitons, which typically possess short diffusion lengths and lifetimes[26,27]. Additionally, a similar density gradient characteristic of diffusion doping was found in the XPS analysis, as shown in Figs 3b and 3c. The characteristic XPS peaks of $MoS_2$ are known to shift towards higher binding energies when additional electrons are introduced by BV molecules, indicative of the $MoS_2$ Fermi level moving towards the conduction band edge[28,29]. Consequently, we observed a peak shift in the directly doped $MoS_2$ film, as illustrated in Fig. 3b. Additionally, XPS spectra were obtained from various positions within the channel area on both directly doped and diffusion doped $MoS_2$ films, as depicted in the upper image of Fig. 3c. While the directly doped $MoS_2$ device exhibited a consistent peak position across all measured positions, the diffusion doped $MoS_2$ one displayed an exponential profile along the channel. It should be noted that no noticeable N-peaks were detected in the diffusion-doped channel area, suggesting that the absence of BV molecules on the $MoS_2$ channel surface provided only additional electrons (Fig. S6). On the contrary, N-peaks are distinctly present in direct-doped $MoS_2$. These results also support the effectiveness of diffusion doping in selectively doping the contact region without affecting the channel area.

To investigate further, we employed scanning Kelvin probe microscopy (SKPM) to measure surface potential profiles modulated by doping in the $MoS_2$ channel. Optical images (Figs. 3e-g) reveal the selective deposition of BV molecules on the CVD $MoS_2$ film using inkjet printing with different dopant concentrations ranging from 1 to 20 mg/mL. A notable increase in the average height of the deposited BV molecules is observed with higher BV concentrations. During the SKPM measurements, the contact potential difference voltage ($V_{CPD}$) was analysed. An increase in $V_{CPD}$ corresponds to a reduction in the work function, which results from an elevated electron concentration in *n*-type $MoS_2$. Remarkably, $V_{CPD}$ decreases consistently along the channel direction, regardless of the BV concentration, as depicted in Figs. 3e-g The $V_{CPD}$ profiles follow an exponentially decreasing curve, signifying a concomitant exponential decline in electron concentration along the channel direction and affirming the diffusion of transferred electrons. Specifically, this exponential decay along the channel aligns with our DFT calculations and XPS results (see Supplementary Information Section 3). Furthermore, we extracted the diffusion length ($L_N$) values using the following electron concentration profile:

$$n(x) = n_0 + n_0(e^{\frac{qV_A}{kT}} - 1)e^{-\frac{x}{L_N}} \qquad (1)$$

where $n_0$ is the initial electron concentration, $q$ is the electron charge, $V_A$ is the applied voltage, $k$ is the Boltzmann's constant, and $T$ is the temperature. The extracted average diffusion lengths were 24.56 µm in Fig. 3e, showing consistency across various dopant concentrations, being an intrinsic physical parameter of $MoS_2$, and exhibited values similar to those of previous theoretical studies[25].

We fabricated an array of diffusion-doped $MoS_2$ FETs using a fixed diffusion doping concentration of 20 mg/mL at the contact region, alongside pristine $MoS_2$ FETs with varying channel lengths. Subsequently, we extracted the total resistance of each device. While the channel resistance of pristine $MoS_2$ FETs increases linearly with channel length (Fig. 4a), their diffusion-doped counterpart exhibits an exponential increase in resistance with increasing channel length (Fig. 4b). This phenomenon can be elucidated by the exponential distribution of diffused electron concentration along the channel. As depicted in Fig. 4c, while pristine FETs without doping maintain a constant resistivity across the channel with a constant carrier density, the carrier concentration in diffusion-doped devices displays an exponential decay profile along the channel, resulting in a corresponding exponential distribution in conductivity. In this context, the total resistance of a diffusion-doped device should exponentially increase as the channel length increases linearly, whereas that of a pristine device should rise linearly. Therefore, given that electron diffusion is predominantly determined by the disparity in electron density between the highly doped contact and channel areas, the diffusion effect can be regulated by varying the electron density in the channel. Diffusion is primarily contingent upon the disparity in charge density, specifically, the difference in charge concentration between the channel and doped contact area, according to the following equation:

$$\frac{\partial^2 \Delta n(x)}{\partial x^2} = \frac{\Delta n}{{L_N}^2} \qquad (2)$$

Consequently, the diffusion doping behaviour will be limited as the difference in charge density, $\frac{dn(x)}{dx}$, diminishes. To validate this hypothesis, a high positive gate voltage was applied to facilitate electrostatic doping in the channel area, thereby increasing the electron concentration in the channel and suppressing diffusion from the doped contact area. As expected, when the channel was electrostatically doped under a high positive gate voltage, the resistance of the diffusion-doped FETs increased linearly, exhibiting significantly different results compared to that without electrostatic doping across the channel length (Figs. 4d-e). As illustrated in Fig. 4f, as the carrier concentration increases owing to electrostatic doping, the carrier density gradient and diffusion doping effect become negligible.

To maximise the potential of semiconducting large-area monolayer TMDCs, it is critical to modify the carrier concentrations to enhance the charge carrier mobility by modulating the screening of charged impurity potentials. We conducted a comparative analysis of charged impurity-limited mobilities and phonon-limited mobilities in pristine, directly doped, and diffusion-doped $MoS_2$ FETs across temperatures ranging from 300 to 10 K. Given that both charged impurity scattering and phonon scattering are predominant factors in reducing the mobility of 2D TMDCs, isolating the effect of phonon scattering by lowering the measurement temperature is crucial for a focused investigation on the impact of charged impurities. Quantitatively, we evaluated the charged impurity-limited mobility ($\mu_{imp}$) at 300 K and 10 K using the theoretical model proposed by Ong and Fischetti, with detailed discussions available in Supplementary Section 6[18,23,30]. For a comprehensive comparison, we also measured the field-effect mobility as a function of carrier concentration by adjusting the dopant concentration. The $\mu_{imp}$ curves at 300 K and 10 K, plotted for various impurity concentrations, are shown in Fig. 5a and b. At 300 K, an appreciable increase in mobility is observed with higher BV dopant concentrations through diffusion doping, despite potential phonon scattering impacts on charge transport. Conversely, no clear trend can be found for the directly doped $MoS_2$ FETs, aside from an increase in charged impurity concentration. Notably, enhanced field-effect mobilities of the diffusion-doped $MoS_2$

FETs originate from reduced charged impurity concentrations, highlighting the effectiveness of diffusion doping in augmenting carrier concentration within the 2D large-area monolayer $MoS_2$ channel, as depicted in Fig. 5a. At 10 K, where charged impurity scattering predominates over phonon scattering, the diffusion-doped $MoS_2$ FETs shows significantly higher mobility values at similar carrier concentrations, but with lower charged impurity concentrations. These findings represent unprecedented advancements over the outcomes of conventional direct SCTD in 2D TMDCs from previous studies.

In summary, we developed a novel doping technique that modulates the electrical properties of CVD-grown monolayer $MoS_2$. This technique effectively overcomes the challenges posed by charged impurities inherent in direct SCTD. We achieved this by selectively depositing BV molecules on the contact region using inkjet printing. This method facilitates the diffusion of donated carriers to the $MoS_2$ channel region without introducing charged impurities. Our theoretical analysis corroborates that stacked BV molecules can transfer electrons more efficiently owing to their larger DOS and higher $E_F$. We conducted a comprehensive quantitative investigation into the diffusion of excess electrons from the directly doped contact region, enriched with high-density BV molecules, to the channel and their doping effects. This was accomplished by employing electrical characterisations, AFM, XPS, and DFT calculations. The impact of charged impurity scattering was meticulously explored through temperature-dependent electrical characterisations. The results indicated that our proposed diffusion doping method can significantly suppress the effects of charged impurities, leading to a considerable improvement in electron mobility. We contend that the diffusion-doping approach effectively resolves a longstanding bottleneck in SCTD and fully exploits the potential of large-area 2D TMDCs by optimising their electrical performance.

# Methods

## Device fabrication

Poly(methyl methacrylate) (PMMA) was spin-coated as a sacrificial layer onto a continuous CVD-grown monolayer $MoS_2$ film on a $SiO_2$/heavily doped silicon substrate(Six-Carbon Corp). This was followed by a hard bake on a hot plate at 180 ℃ for 90 seconds. Subsequently, thermal releasing supporting tape (Graphene Square) was attached directly to the PMMA-coated $MoS_2$ film. The assembly was immersed in a 60 ℃ potassium hydroxide (KOH) solution to etch away the $SiO_2$ layer, then rinsed with deionized water and dried. The PMMA-coated $MoS_2$ film was transferred and dried in a vacuum oven for 12 hours. Following the thermal treatment at 120 ℃ to remove the supporting thermal release tape, the PMMA sacrificial layer was dissolved in acetone for 10 minutes. BV doping ink was then selectively printed on the contact area followed by an annealing step at 120 ℃ for one hour in a vacuum oven to eliminate any residual solvents (Omnijet 300, Unijet). While printing, a double-pulse waveform was employed to enhance the stability of the BV droplets and the plate temperature was maintained at 60℃ (1pL, Dimatix cartridge). Ag ink (DGP 40LT-15C, ANP Co. Ltd.) for electrical contacts were directly inkjet-printed onto the $MoS_2$.

## Device characterisations

The electrical characterisations were conducted in a temperature variable probe station (M6VC, MS-Tech) with a semiconductor parameter analyser (B1500A, Keysight) under the vacuum condition at room temperature. Temperature-dependent I-V measurements were conducted using a cryostat system (CS204*I-FMX-12, Advanced Research Systems).

XPS spectra were obtained using an X-ray beam with a 20 µm diameter, a power of 4.5 W, and an acceleration voltage of 15 kV (PHI Quantera-II, Ulvac-PHI). For the cross-sectional TEM (TITAN 80-300, ThermoFisher Scientific) analysis, samples were prepared by using a focused ion beam (FIB) system (G4-HX). BV deposited area of $MoS_2$ was imaged by SEM (JSM-6510, JEOL) and the surface energy of $MoS_2$ was measured *via* AFM (NX10, Park Systems).

**DFT calculations**

The electronic structures of the $MoS_2$/BV were calculated at the density functional theory (DFT) level of theory using Perdew-Burke-Ernzerhof (PBE) generalized gradient approximation, as implemented in the Vienna Ab Initio Simulation Package (VASP).

## References

1. Manzeli, S., Ovchinnikov, D., Pasquier, D., Yazyev, O. V. & Kis, A. 2D transition metal dichalcogenides. *Nat. Rev. Mater.* **2**, 1–15 (2017).
2. Wang, Q. H., Kalantar-Zadeh, K., Kis, A., Coleman, J. N. & Strano, M. S. Electronics and optoelectronics of two-dimensional transition metal dichalcogenides. *Nat. Nanotechnol.* **7**, 699–712 (2012).
3. Chhowalla, M., Jena, D. & Zhang, H. Two-dimensional semiconductors for transistors. *Nat. Rev. Mater.* **1**, 1–15 (2016).
4. Beck, M. E. & Hersam, M. C. Emerging opportunities for electrostatic control in atomically thin devices. *ACS Nano* **14**, 6498–6518 (2020).
5. Amani, M. *et al.* Near-unity photoluminescence quantum yield in $MoS_2$, *Science* **350**, 1065–1068 (2015).
6. Zhang, T. *et al.* Universal in situ substitutional doping of transition metal dichalcogenides by liquid-phase precursor-assisted synthesis. *ACS Nano* **14**, 4326–4335 (2020).
7. Hu, Z. *et al.* Two-dimensional transition metal dichalcogenides: interface and defect engineering. *Chem. Soc. Rev.* **47**, 3100–3128 (2018).
8. Ristein, J. Surface transfer doping of semiconductors. *Science* **313**, 1057–1058 (2006).
9. Zhao, Y. *et al.* Doping, contact and interface engineering of two-dimensional layered transition metal dichalcogenides transistors. *Adv. Funct. Mater.* **27**, 1603484 (2017).
10. Mouri, S., Miyauchi, Y. & Matsuda, K. Tunable Photoluminescence of monolayer $MoS_2$ via chemical doping. *Nano Lett.* **13**, 5944–5948 (2013).
11. Fang, H. *et al.* High-performance single layered $WSe_2$ p-FETs with chemically doped contacts. *Nano Lett.* **12**, 3788–3792 (2012).

12. Yang, L. *et al.* Chloride molecular doping technique on 2D materials: $WS_2$ and $MoS_2$. *Nano Lett.* **14**, 6275–6280 (2014).

13. Pandey, S. K. *et al.* Controlled p-type substitutional doping in large-area monolayer $WSe_2$ crystals grown by chemical vapor deposition. *Nanoscale* **10**, 21374–21385 (2018).

14. Zhang, S. *et al.* Controllable, wide-ranging n-doping and p-doping of monolayer group 6 transition-metal disulfides and diselenides. *Adv. Mater.* **30**, 1802991 (2018).

15. Li, S.-L., Tsukagoshi, K., Orgiu, E. & Samorì, P. Charge transport and mobility engineering in two-dimensional transition metal chalcogenide semiconductors. *Chem. Soc. Rev.* **45**, 118–151 (2015).

16. Kiriya, D., Tosun, M., Zhao, P., Kang, J. S. & Javey, A. Air-stable surface charge transfer doping of $MoS_2$ by benzyl viologen. *J. Am. Chem. Soc.* **136**, 7853–7856 (2014).

17. Liu, H., Liu, Y. & Zhu, D. Chemical doping of graphene. *J. Mater. Chem.* **21**, 3335–3345 (2011).

18. Kim, J.-K. *et al.* Molecular dopant-dependent charge transport in surface-charge-transfer-doped tungsten diselenide field effect transistors. *Adv. Mater.* **33**, 2101598 (2021).

19. Shi, W. *et al.* Reversible writing of high-mobility and high-carrier-density doping patterns in two-dimensional van der Waals heterostructures. *Nat. Electron.* **3**, 99–105 (2020).

20. Ma, N. & Jena, D. Charge scattering and mobility in atomically thin semiconductors. *Phys. Rev. X* **4**, 011043 (2014).

21. Li, S.-L. *et al.* Thickness-dependent interfacial coulomb scattering in atomically thin field-effect transistors. *Nano Lett.* **13**, 3546–3552 (2013).

22. Lee, D. *et al.* Remote modulation doping in van der Waals heterostructure transistors. *Nat. Electron.* **4**, 664–670 (2021).

23. Jang, J. *et al.* Reduced dopant-induced scattering in remote charge-transfer-doped $MoS_2$ field-effect transistors. *Sci. Adv.* **8**, eabn3181 (2022).

24. Jeong, I. *et al.* Tailoring the electrical characteristics of $MoS_2$ FETs through controllable surface charge transfer doping using selective inkjet printing. *ACS Nano* **16**, 6215–6223 (2022).

25. Chaves, F. A., Feijoo, P. C. & Jiménez, D. 2D pn junctions driven out-of-equilibrium. *Nanoscale Adv.* **2**, 3252–3262 (2020).

26. Massicotte, M. *et al.* Picosecond photoresponse in van der Waals heterostructures. *Nat. Nanotechnol.* **11**, 42–46 (2016).

27. Yu, W. J. *et al.* Unusually efficient photocurrent extraction in monolayer van der Waals heterostructure by tunnelling through discretized barriers. *Nat. Commun.* **7**, 13278 (2016).

28. Chamlagain, B., Withanage, S. S., Johnston, A. C. & Khondaker, S. I. Scalable lateral heterojunction by chemical doping of 2D TMD thin films. *Sci. Rep.* **10**, 12970 (2020).

29. Qu, D. *et al.* Carrier-type modulation and mobility improvement of thin $MoTe_2$. *Adv. Mater.* **29**, 1606433 (2017).

30. Ong, Z.-Y. & Fischetti, M. V. Mobility enhancement and temperature dependence in top-gated single-layer $MoS_2$. *Phys. Rev. B* **88**, 165316 (2013).

## Author contributions

K.C. and S.C. conceptualised and designed the experiments. The device fabrication and measurements were conducted by I.J., J.Y., and J.T.. D.C. performed the DFT simulation and D.H.K. obtained the high-quality TEM images. J.K.K. conducted the calculation of charged impurity concentrations.  T.L, K.C, and S.C. supervised the project. All authors discussed the results and commented on the manuscript, which was written by K.C. and S.C.

## Acknowledgements

The authors appreciate the financial support from the National Research Foundation of Korea (NRF) grant funded by the government of the Republic of Korea (the Ministry of Science and ICT) (No NRF-2023R1A2C2003247, 2021R1C1C2091728, 2021R1A2C3004783). This work was also supported by the National Research Foundation of Korea (NRF) grant funded by the Korea government (MSIT) (No. RS-2024-00416978)

## Additional information

Supplementary information is available for this paper here (TBD). Reprints and permissions information is available at www.nature.com/reprints. Correspondence and requests for materials should be addressed to K.C. or S.C.

## Competing interests

The authors declare no competing interests.

## Data availability

The data that support the plots within this study are available from the corresponding authors upon reasonable request.

## Figure legends

**Fig. 1. Principles of electron donation in $MoS_2$ *via* direct and diffusion doping methods a**, Schematic illustrations of the direct doping process. After the donation of electrons, molecular dopants, which can behave as charged impurities, remain at the surface of the channel. **b**, Schematic illustrations of the diffusion doping process. Molecular dopants were deposited on the contact regions, not on the channel region. **c**, Density of States (DOS) of pristine $MoS_2$ and BV-deposited $MoS_2$ with different numbers of BV layers, calculated using Density Functional Theory (DFT). Red lines represent the Fermi level. **d**, Schematic image of a diffusion-doped $MoS_2$ FET. BV molecules are located between metal electrodes and $MoS_2$. Transferred electrons from BV molecules diffuse into the channel area due to the carrier concentration difference.

**Fig. 2. Electrical characterisations of diffusion doped $MoS_2$ FETs** Transfer curves of diffusion doped $MoS_2$ FETs with different BV doping concentrations; **a**, 1 mg/mL, **b,** 5 mg/mL and **c,** 20 mg/mL, respectively. Output curves of diffusion doped $MoS_2$ FETs at $V_g$ = 60 V with different BV doping concentrations; **d**, 1 mg/mL, **e,** 5 mg/mL and **f,** 20 mg/mL, respectively. Insets are photographs of the BV dopant solution for each concentration. **g**, Schematic illustration of series resistance in the $MoS_2$ FETs before and after diffusion doping.

**Fig. 3. Experimental evidence of diffusion doping a**, Schematic illustration of the energy band across the $MoS_2$ channel after diffusion doping. As electrons diffuse into the channel area, potential energy builds up, preventing further diffusion. **b**, XPS spectra of pristine $MoS_2$ film and BV-deposited $MoS_2$ (doped) film. Black dashed lines indicate peak positions of pristine $MoS_2$ and red lines indicate those of doped $MoS_2$. **c**, Mo $3d_{3/2}$ peak positions of direct doped and diffusion

doped $MoS_2$ FETs along the channel. The upper panel is the SEM image of a $MoS_2$ FET and red circles are points where XPS data was measured. White regions are metal electrodes. **e-g**, OM images, AFM images, height profiles, SKPM images, and energy profiles of diffusion-doped $MoS_2$ films with different doping concentrations: e) 1 mg/mL, f) 5 mg/mL, and g) 20 mg/mL, respectively. The left part of the white dashed line indicates where the BV dopant molecules are selectively deposited.

**Fig. 4. Diffusion doping efficiency depending on channel conductivity and channel length** Total resistance of **a**, pristine $MoS_2$ FETs **b**, diffusion doped $MoS_2$ FETs without electrostatic doping, varying channel length. While the total resistance of pristine devices increases linearly, those of diffusion doped devices increase exponentially as channel length increases. **c**, Energy band schematic of a pristine and a diffusion doped device without gating. Dashed lines indicate carrier concentration and resistivity of the pristine sample and solid lines indicate those of the diffusion doped sample. Total resistance of **d**, pristine $MoS_2$ FETs **e**, diffusion doped $MoS_2$ FETs at $V_g$ = 60 V, varying channel length. Due to heavy electrostatic doping, both pristine and diffusion doped devices show a linear increase in total resistance as the channel length increases. **f**, Energy band schematic of a pristine and a diffusion doped device with electrostatic doping as well. Dashed lines indicate carrier concentration and resistivity of the pristine sample and solid lines indicate those of the diffusion doped sample.

**Fig. 5. Field-effect mobility of $MoS_2$ FETs depending on the doping strategies a,** Carrier mobility values of pristine, direct doped, and diffusion doped devices at 300 K with various doping concentrations. Solid lines indicate the calculated mobility values depending on carrier

concentrations at 300 K. Carrier mobility decreases as the charged impurity concentration increases. **b**, Carrier mobility values of pristine, direct doped, and diffusion doped devices at 10 K with various doping concentrations. Solid lines indicate the calculated mobility values depending on carrier concentrations at 10 K.

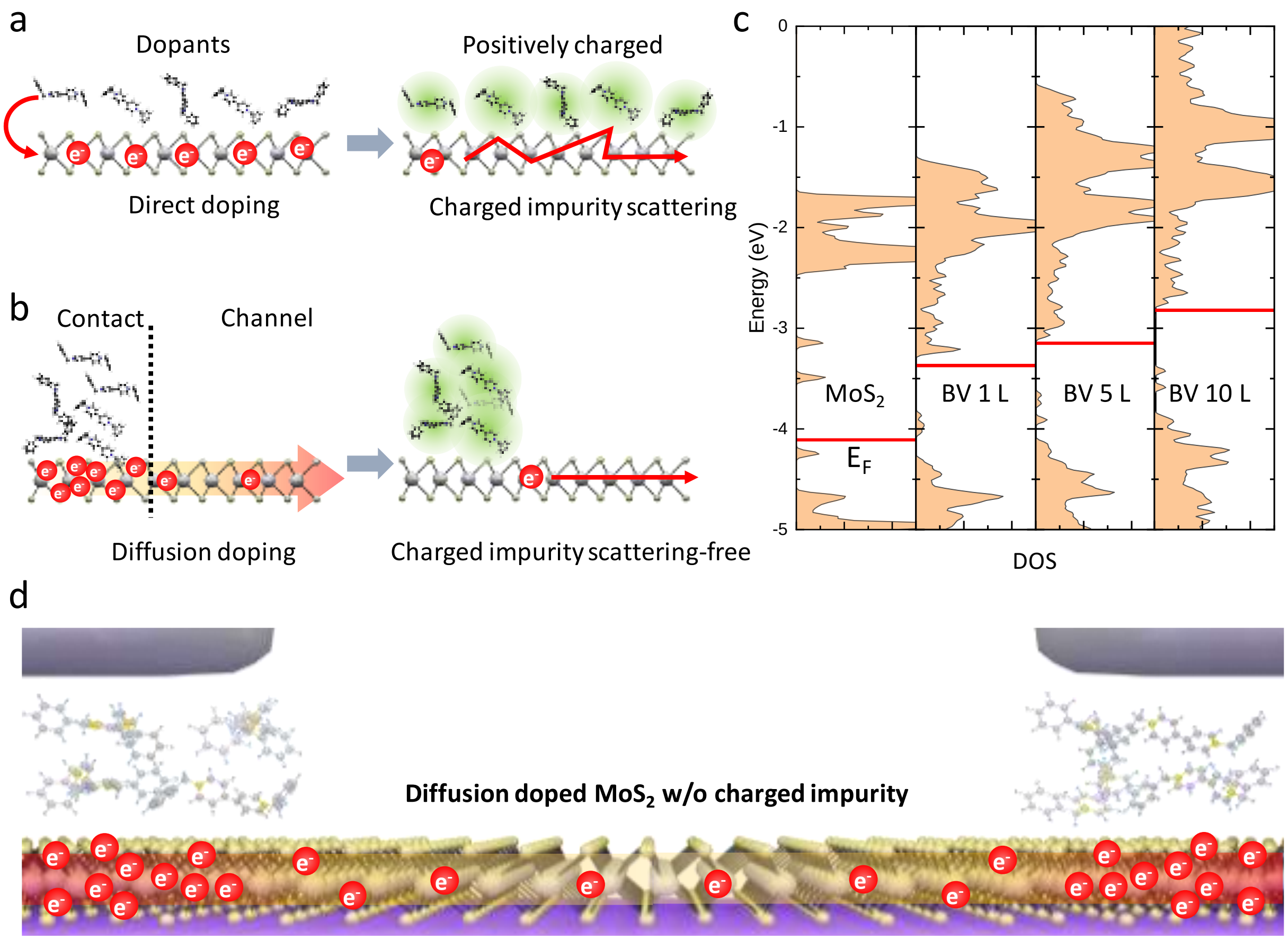

Figure 1. Jeong *et al*.

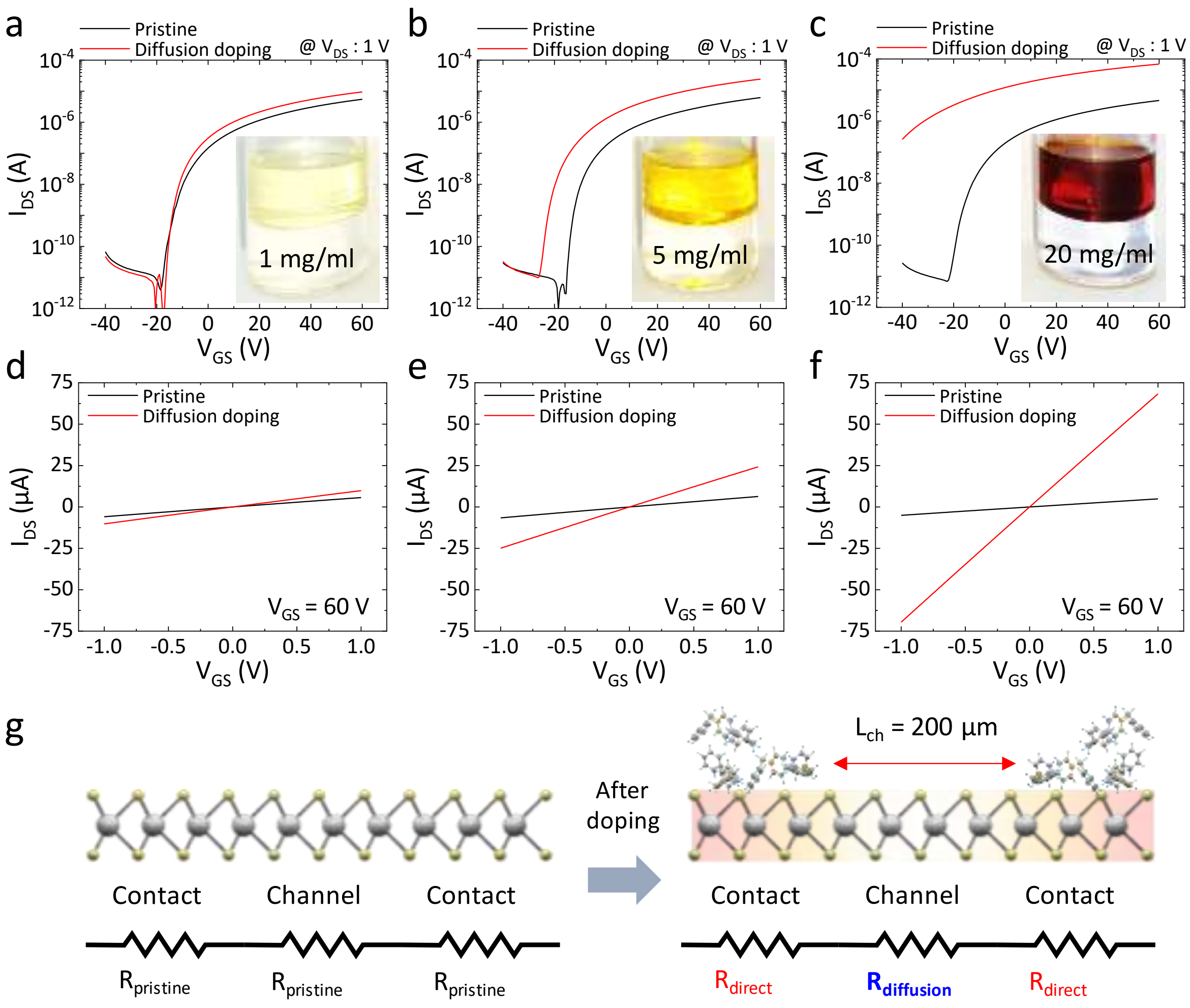

Figure 2. Jeong *et al.*

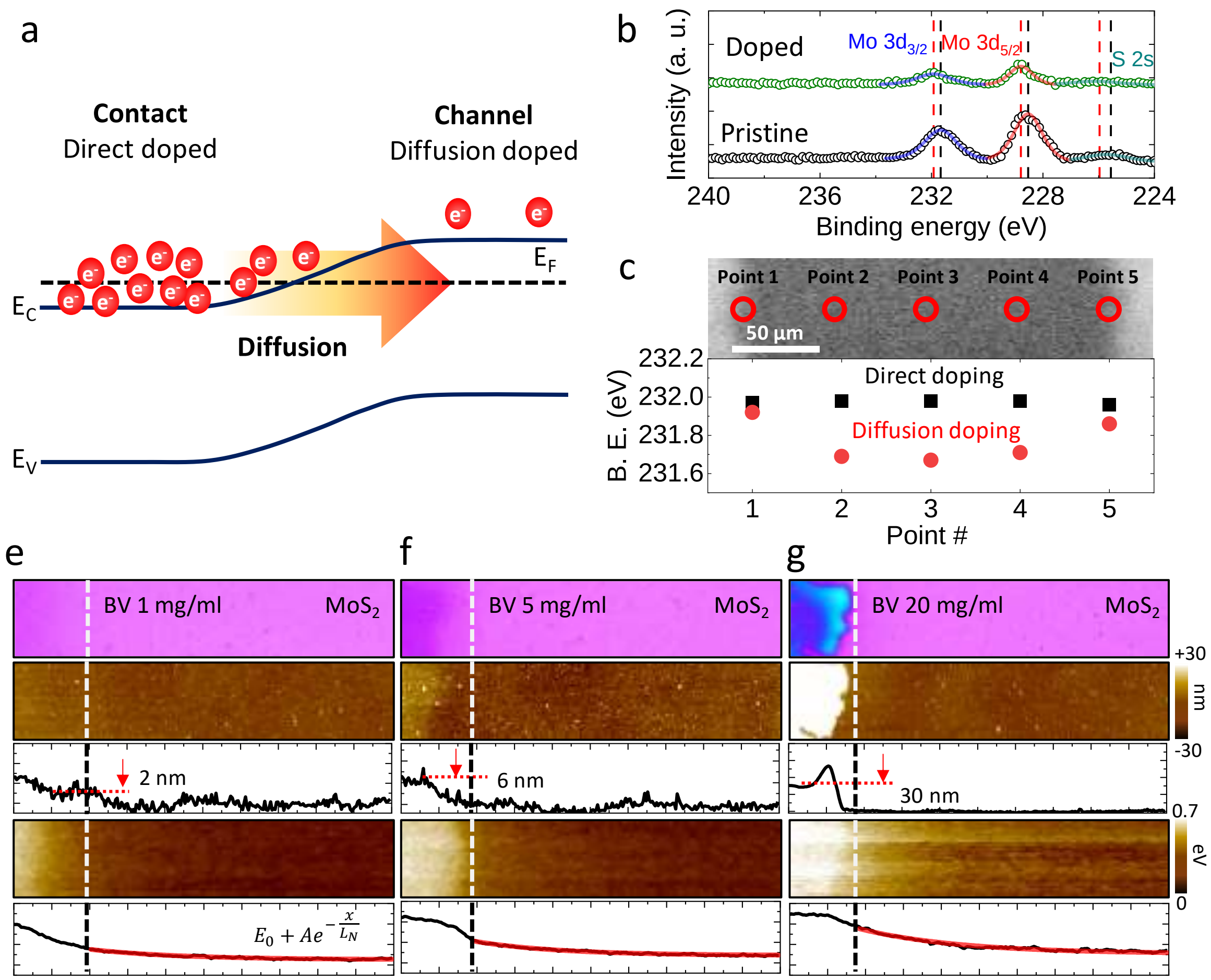

Figure 3. Jeong *et al.*

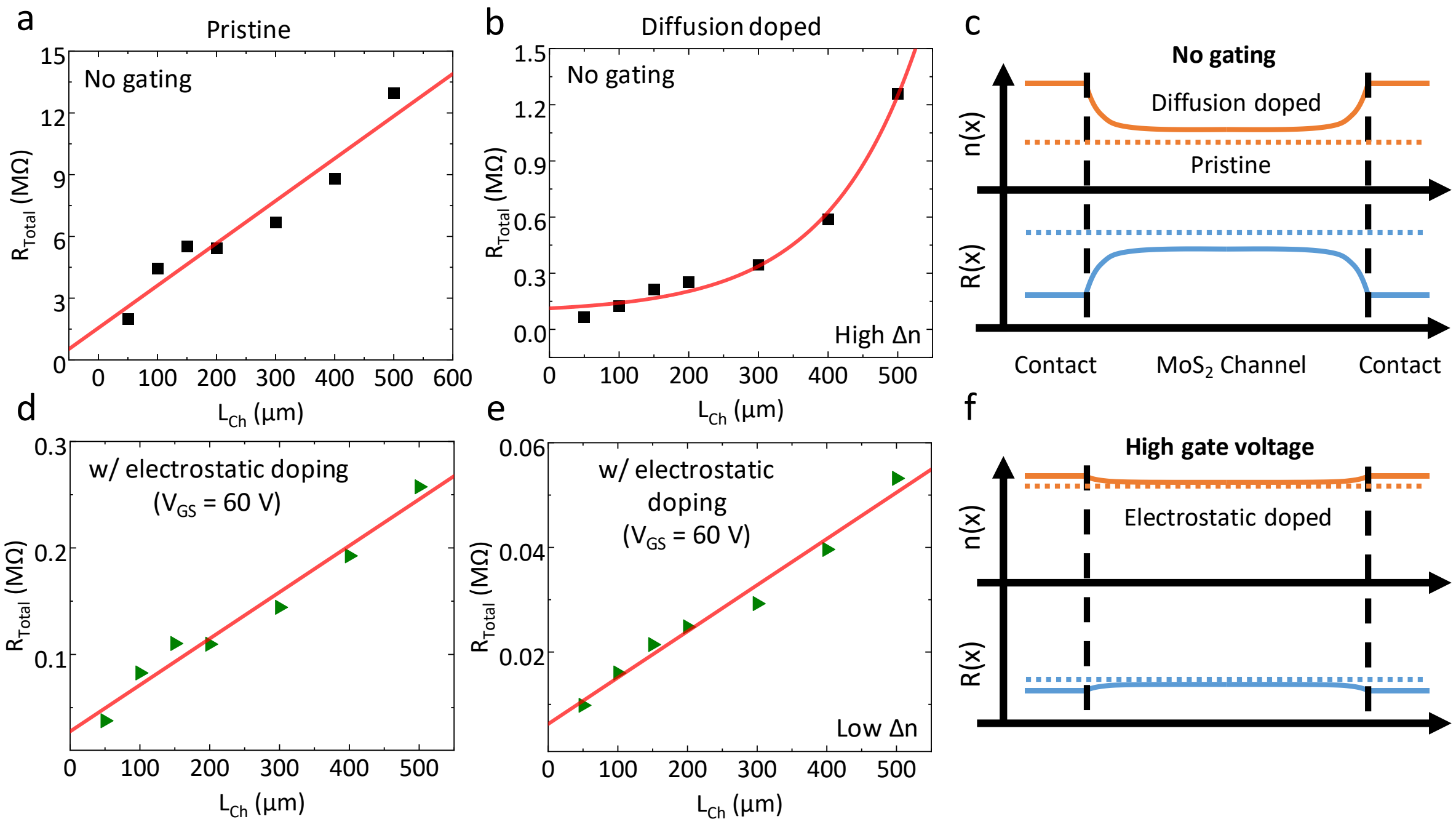

Figure 4. Jeong *et al*.

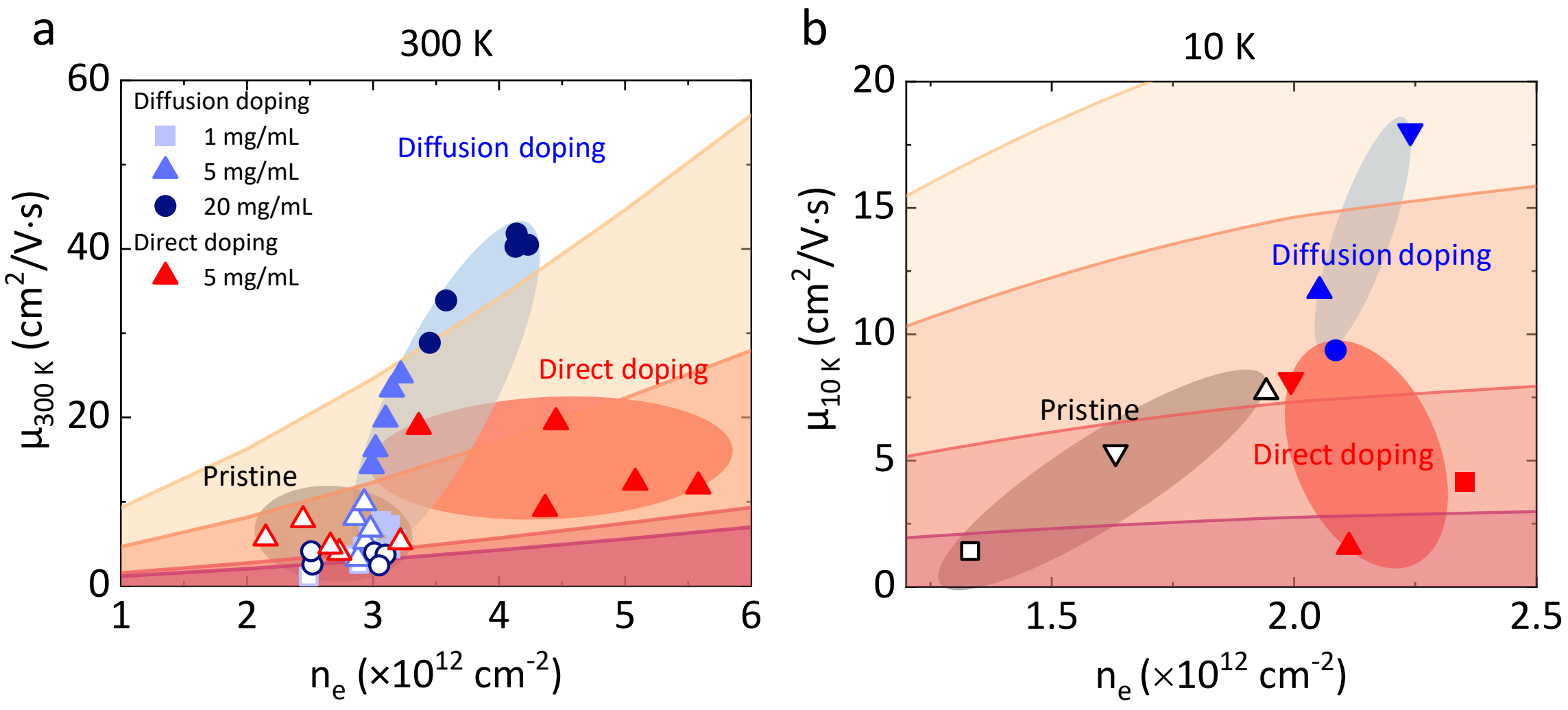

Figure 5. Jeong *et al*.

Supplementary Information for

# Charged-impurity free printing-based diffusion doping in molybdenum disulfide field-effect transistors

Inho Jeong†, Jiwoo Yang†, Juntae Jang†, Daeheum Cho, Deok-Hwang Kwon, Jae-Keun Kim, Takhee Lee*, Kyungjune Cho*, and Seungjun Chung*

Correspondence and requests for materials should be addressed to Takhee Lee (e-mail: tlee@snu.ac.kr), Kyungjune Cho (e-mail: kcho@kist.re.kr) and Seungjun Chung (seungjun@korea.ac.kr).

†These authors contributed equally to this work.

Table of Contents

## Section 1: Device fabrication

### 1-1. BV ink formulation

Benzyl viologen (BV) dichloride powders were solubilized in 5 mL of deionized water and subsequently mixed with 5 mL of a nonpolar solvent to create a biphasic mixture. For the ink, a 4:6 ratio of toluene and cyclohexylbenzene was used to guarantee the good printability. As demonstrated in Fig. S1, this specific ratio was selected due to its superior printability, attributed to optimal rheological parameters. Sodium borohydride powder, procured from Sigma–Aldrich and serving as a catalyst, was then added to the biphasic solution. After 24 hours, the upper layer, enriched with the BV dopant, was extracted using a syringe.

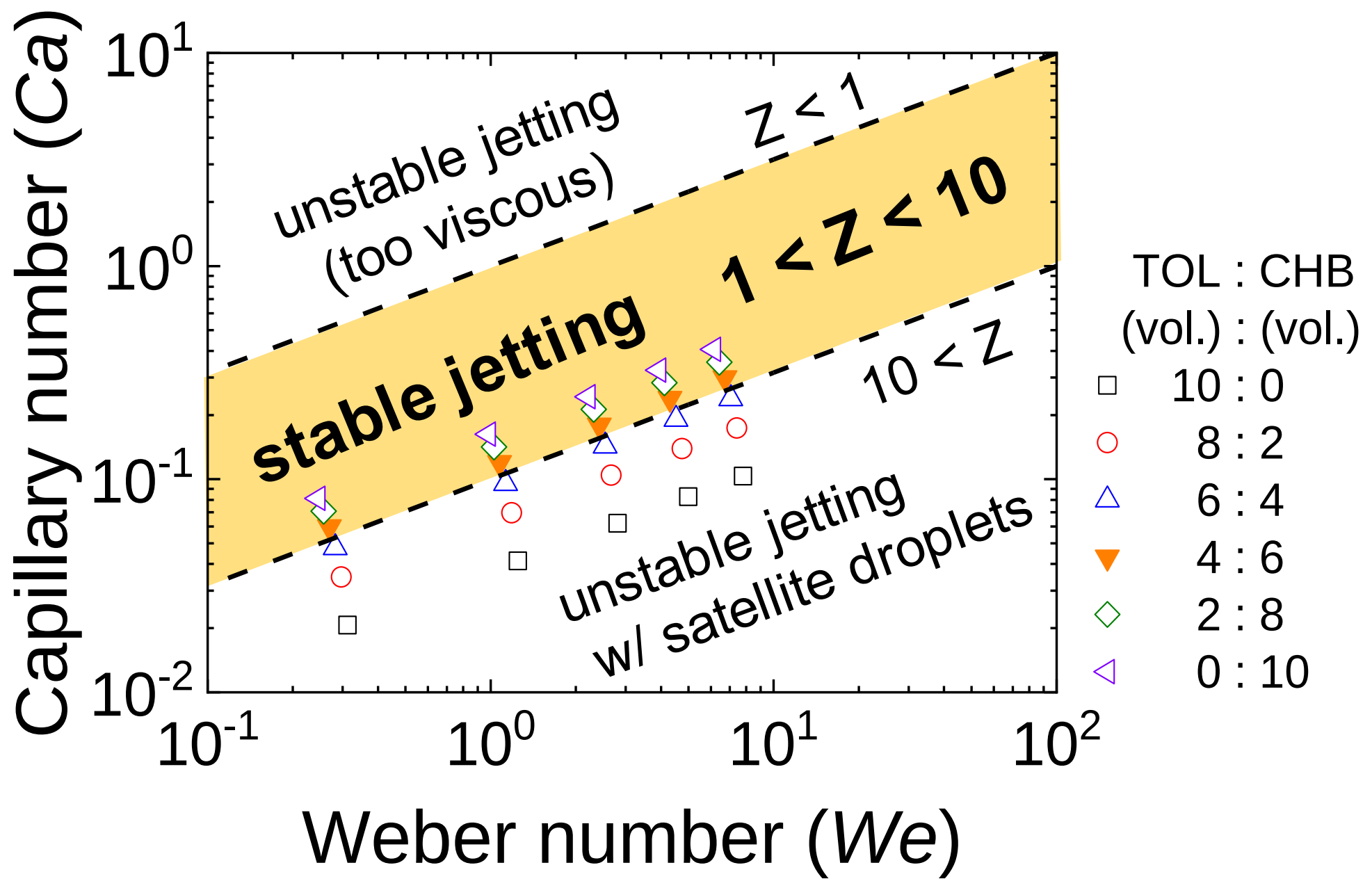

**Fig. S1**. Rheological parameters of BV ink with various TOL:CHB ratios

For the doping process of $MoS_2$ FETs, inkjet printing was employed. Following this, the BV-doped $MoS_2$ FETs underwent an annealing process at 120 ℃ for one hour in a vacuum oven, aimed at removing any remaining solvents. As shown in Fig. S2, the formulated BV inks with different BV concentrations well-formed droplets for the inkjet printing process.

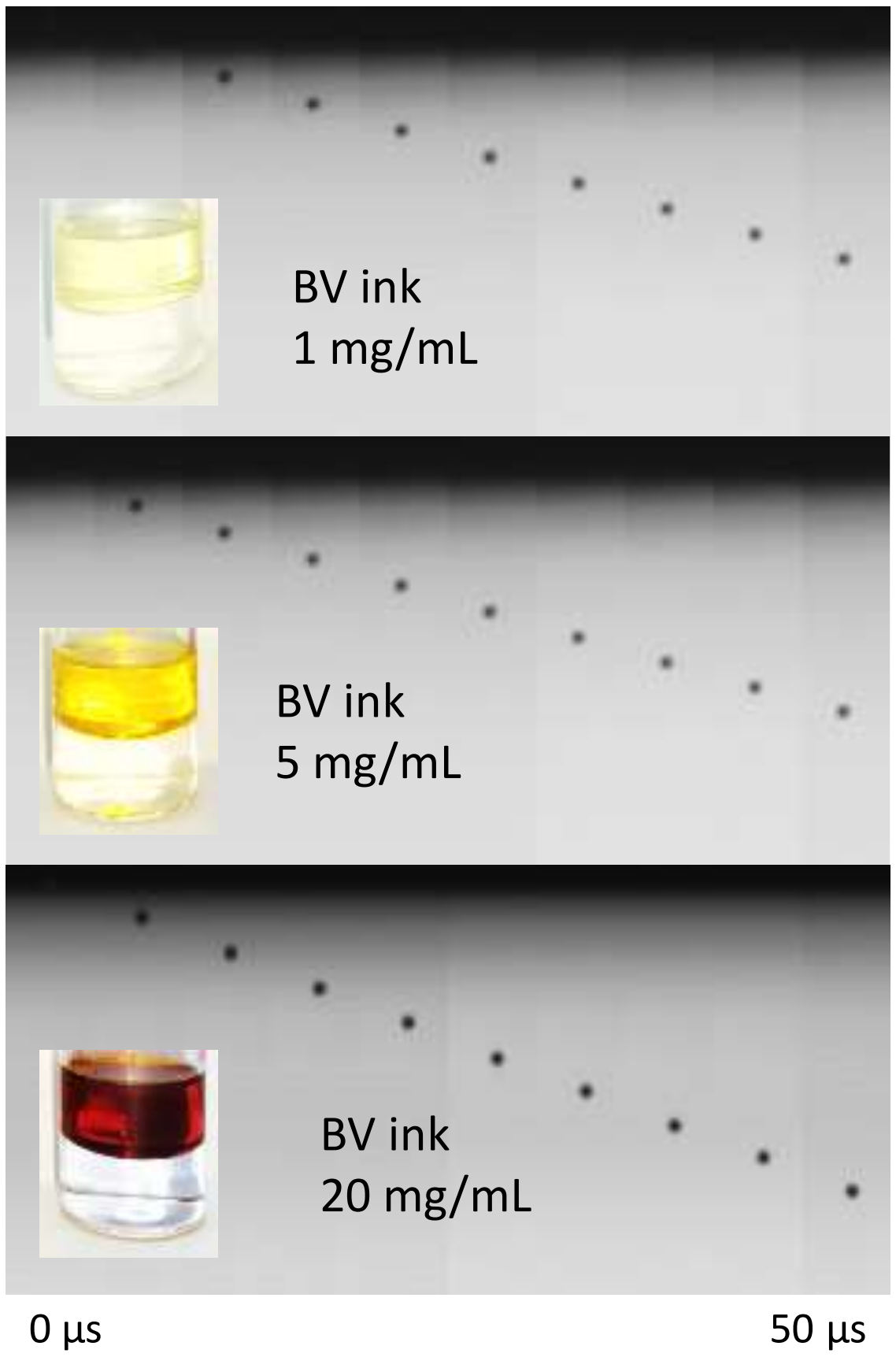

**Fig. S2**. Ink droplet formations with different BV concentrations

**1-2. Detailed fabrication procedures**

Poly(methyl methacrylate) (PMMA) was spin-coated as a sacrificial layer onto a continuous CVD-grown monolayer $MoS_2$ film on a $SiO_2$/heavily doped silicon substrate, supplied by Six-Carbon Corp. This was followed by a hard bake on a hot plate at 180 ℃ for 90 sec. Subsequently, thermal releasing supporting tape, obtained from Graphene Square, was attached directly to the PMMA-coated $MoS_2$ film. The assembly was immersed in a 60 ℃ potassium hydroxide (KOH) solution to etch away the $SiO_2$ layer, then rinsed with deionized water and dried. The PMMA-coated $MoS_2$ film was transferred to the target substrate and dried in a vacuum oven for 12 hours. Following the thermal treatment at 120 ℃ to remove the supporting thermal release tape, the PMMA sacrificial layer was dissolved in acetone at 60 ℃ for 10 min, exposing the monolayer $MoS_2$ films. BV doping ink was then selectively printed on the contact area followed by an annealing step at 120 ℃ for one hour in a vacuum oven to eliminate any residual solvents (Omnijet 300, Unijet). While printing, a double-pulse waveform was employed to enhance the stability of the BV droplets and the plate temperature was maintained at 60℃ (1pL, Dimatix cartridge). Afterward, electrical contacts were directly inkjet-printed onto the $MoS_2$ transferred to the substrate, utilizing nanoparticle-type Ag ink (DGP 40LT-15C, ANP Co. Ltd.). Following the printing of source and drain electrodes with a channel width of 500 µm and a channel length of 100 µm (10pL, Dimatix cartridge), they were annealed at 140 ℃ for one hour on a hot plate. The electrical properties of both pristine and doped $MoS_2$ FETs were characterized using a semiconductor parameter analyser (Agilent B1500A) in vacuum conditions ( Probe-station, MS-Tech)

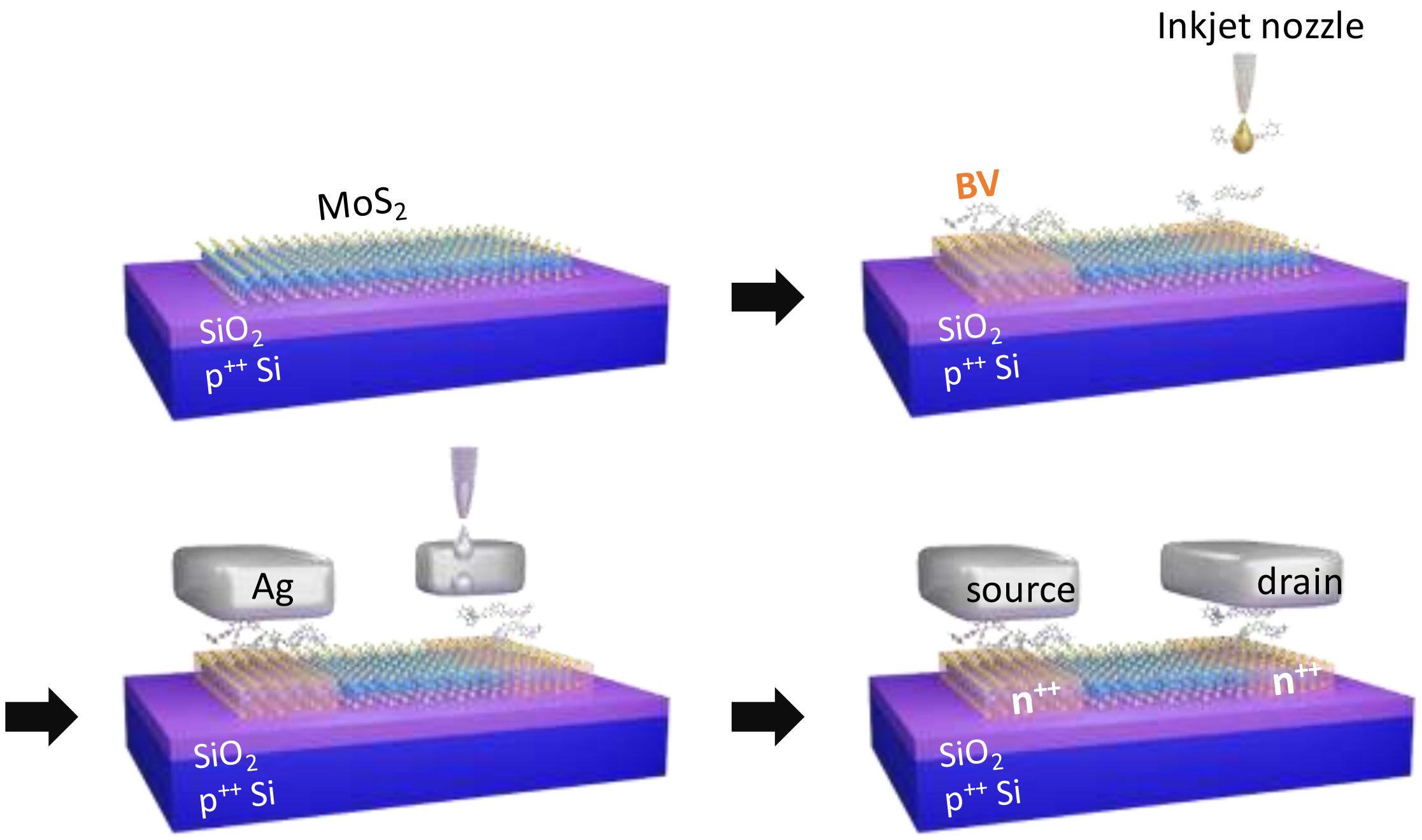

**Fig. S3**. Schematic illustrations of device fabrication process

**1-3. Cross-sectional Energy-filtered transmission electron microscopy (EFTEM) analysis**

Cross-sectional Energy-filtered Transmission Electron Microscopy (EFTEM) analysis was performed to verify the presence of BV molecules situated between the $MoS_2$ and Ag electrode, ensuring their retention following the printing of Ag electrodes. For the cross-sectional TEM analysis, samples were prepared by using a focused ion beam (FIB) system (G4-HX). We observed a distinct carbon signal from the BV molecule deposited on the $MoS_2$ surface, whereas no noticeable carbon layer was detected in pristine $MoS_2$ as shown in Fig. S4 (Titan).

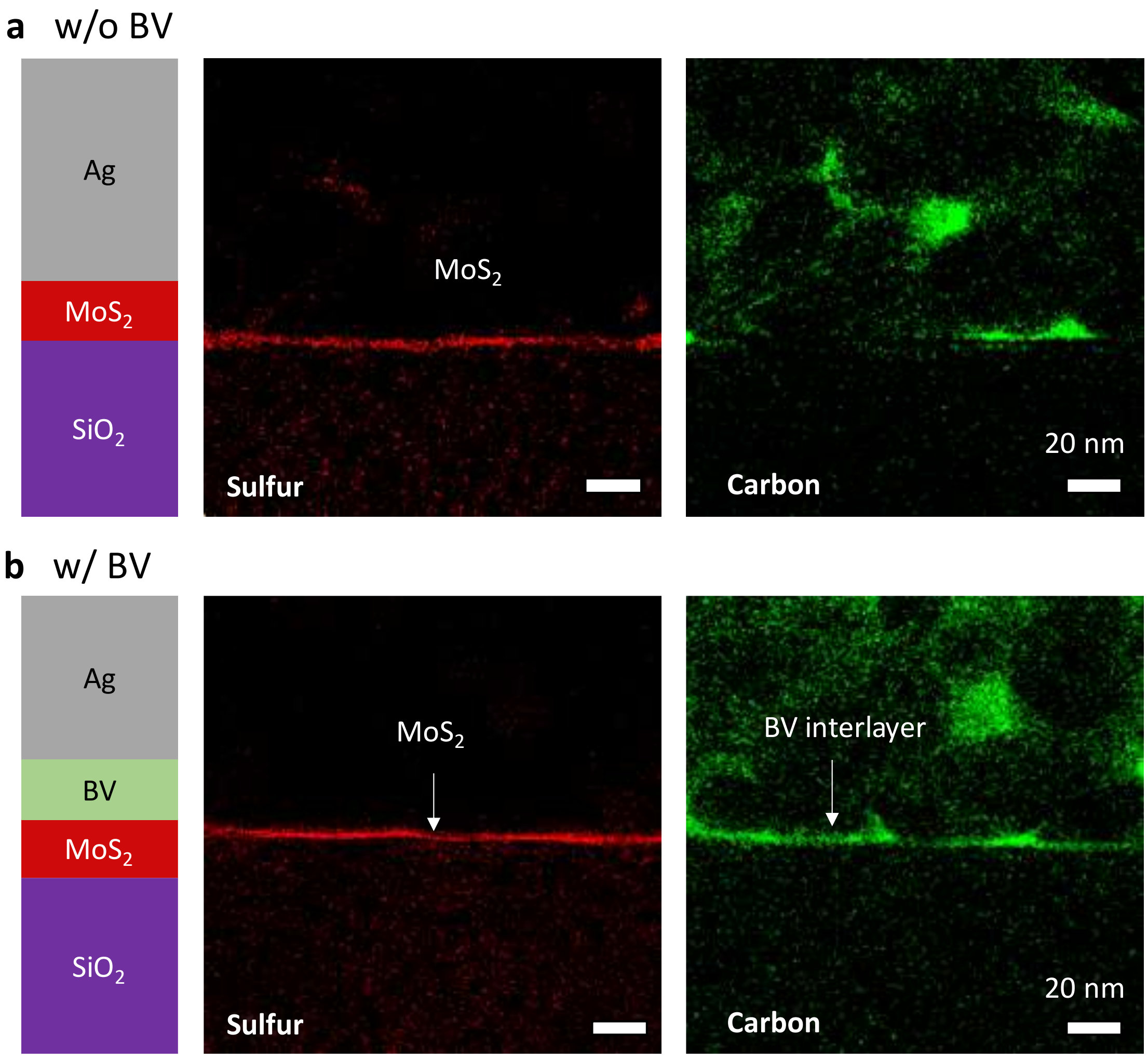

**Fig. S4**. Cross-sectional EFTEM images **a**, without BV and **b**, with BV

**1-4. Selective deposition of BV molecules**

The inkjet printing technique facilitates the selective deposition of BV ink on the CVD-$MoS_2$ layer. As depicted in Fig. S5, deposited BV molecules on the $MoS_2$ surface were observed using SEM (JSM-6510, JEOL). It is important to note that the BV ink did not splash onto the channel areas.

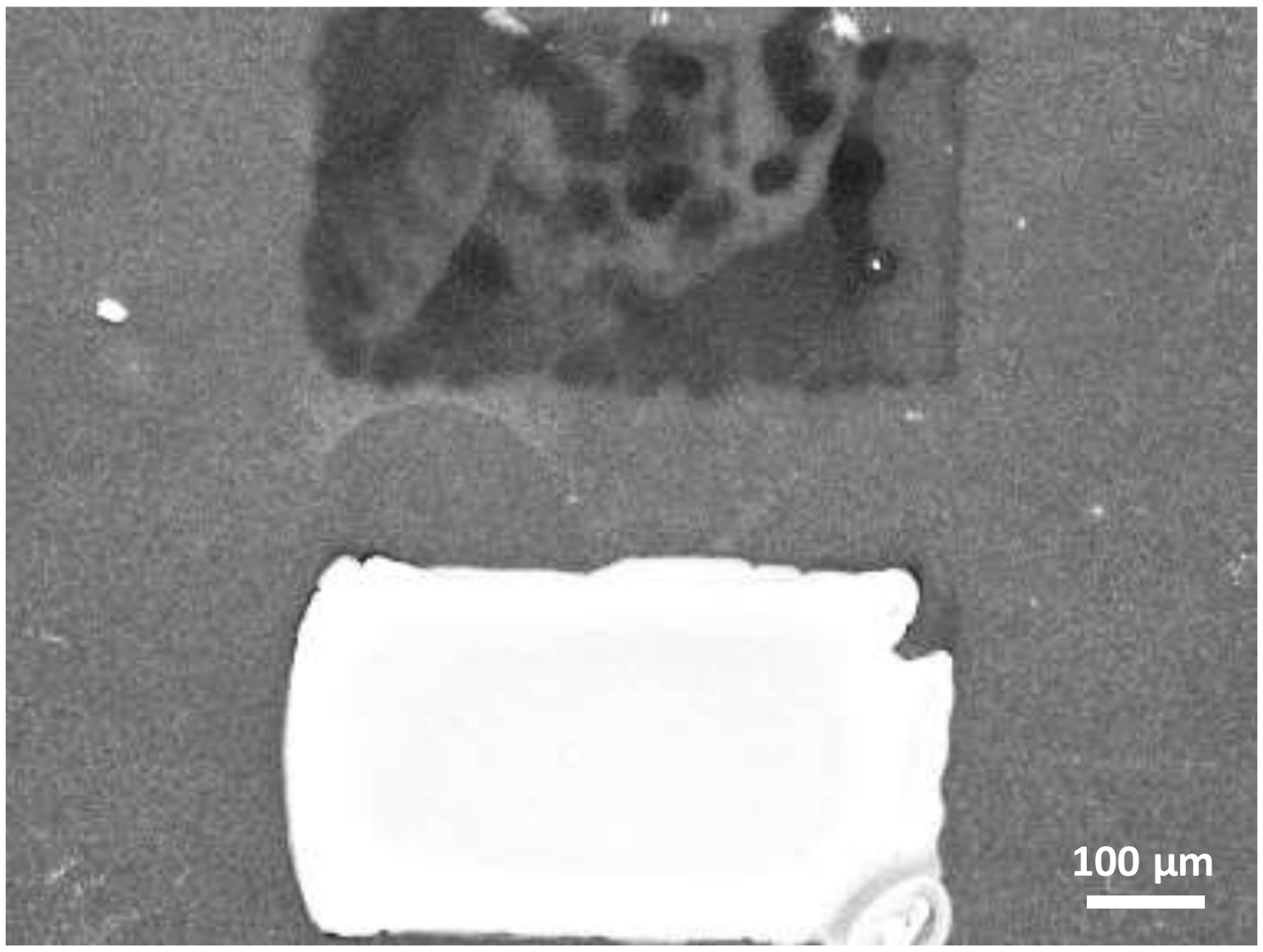

**Fig. S5**. SEM image of a diffusion-doped $MoS_2$ FET. A silver electrode was deposited at the source only. The darker areas indicate BV molecules.

## Section 2: XPS analysis of pristine and doped $MoS_2$

Figs. S6a-c exhibit the XPS spectra of undoped $MoS_2$ (top), directly doped (middle), and diffusion doped $MoS_2$ (bottom) films, respectively. These spectra were obtained using an X-ray beam with a 20 μm diameter, a power of 4.5 W, and an acceleration voltage of 15 kV (PHI Quantera-II, Ulvac-PHI). The spectra reveal distinct photoelectron signals at a binding energy around 399 eV associated with N 1s of BV molecules, observed after the direct introduction of BV; this peak is not present in pristine $MoS_2$. It is noteworthy that in the channel area of the diffusion doped $MoS_2$ film, as shown in Fig. S6c, there is no corresponding BV peak, since no BV molecules are deposited on the channel area.

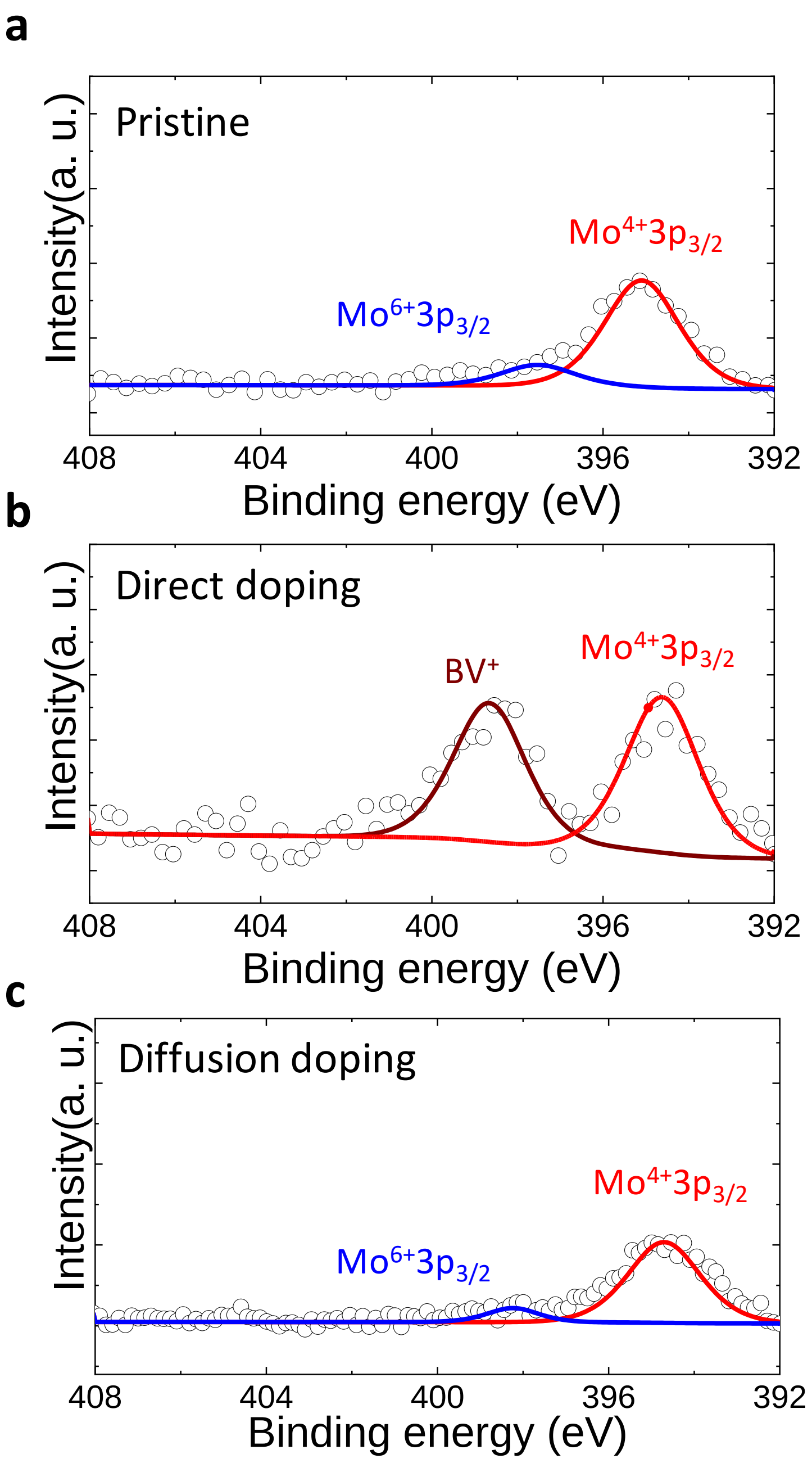

**Fig. S6**. XPS spectrum obtained from the channel area of **a,** pristine, **b,** directly doped, and **c,** diffusion-doped $MoS_2$ FETs.

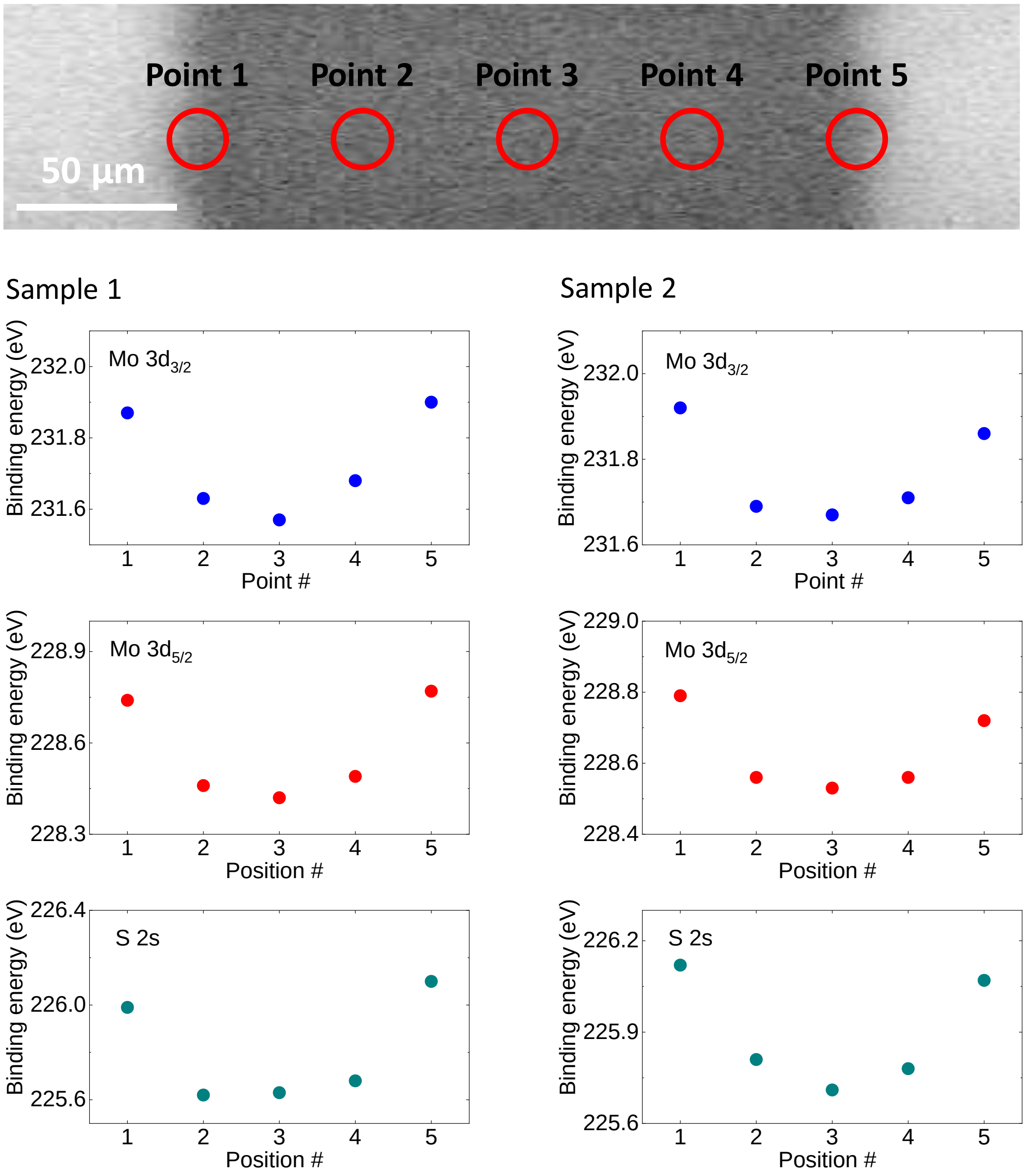

**Fig. S7**. Characteristic peak positions of two diffusion doped $MoS_2$ FETs along the channel. The upper panel is the SEM image of a $MoS_2$ FET and red circles are points where we acquired XPS data. White regions are metal electrodes.

Not only the Mo3d$_{3/2}$ peaks, as mentioned in the main article, but also the Mo3d$_{5/2}$ and S2$_s$ peaks exhibited the same shift tendency across different devices.

## Section 3: DFT calculations

### 3-1. Computational details

The electronic structures of the $MoS_2$/BV and the $WSe_2$/$F_4$-TCNQ were calculated at the density functional theory (DFT) level of theory using Perdew-Burke-Ernzerhof (PBE) generalized gradient approximation, as implemented in the Vienna Ab Initio Simulation Package (VASP)[S1,S2]. The DFT-D3 method with Becke-Johnson damping function was employed to properly describe the dispersion interaction between the $MoS_2$/ BV and $WSe_2$/F4TCNQ[S3]. For all periodic calculations, we used a $\Gamma$-point sampling of the Brilliouin zone for supercells and a 400 eV plane-wave cutoff energy. The vacuum space of 20 Å was applied in the z-direction to eliminate the interactions between the periodic images.

The geometries of BV and F4TCNQ monomers were optimized using the B3LYP/6-31+G* as implemented in the ORCA program package and then further optimized using the VASP program[S4,S5]. The optimal distance between the $WSe_2$/$F_4$-TCNQ were determined by scanning the distance between the $MoS_2$ (or $WSe_2$) and the BV (or $F_4$-TCNQ) while fixing the lattice parameters and the internal coordinates of the $MoS_2$ (or $WSe_2$) slab and the BV (or $F_4$-TCNQ). The optimal distance between the BV (or $F_4$-TCNQ) layers was also determined by the same method. The sulfur vacancies in the $MoS_2$ and $WSe_2$ surfaces were considered in systems.

### 3-2. $MoS_2$/BV

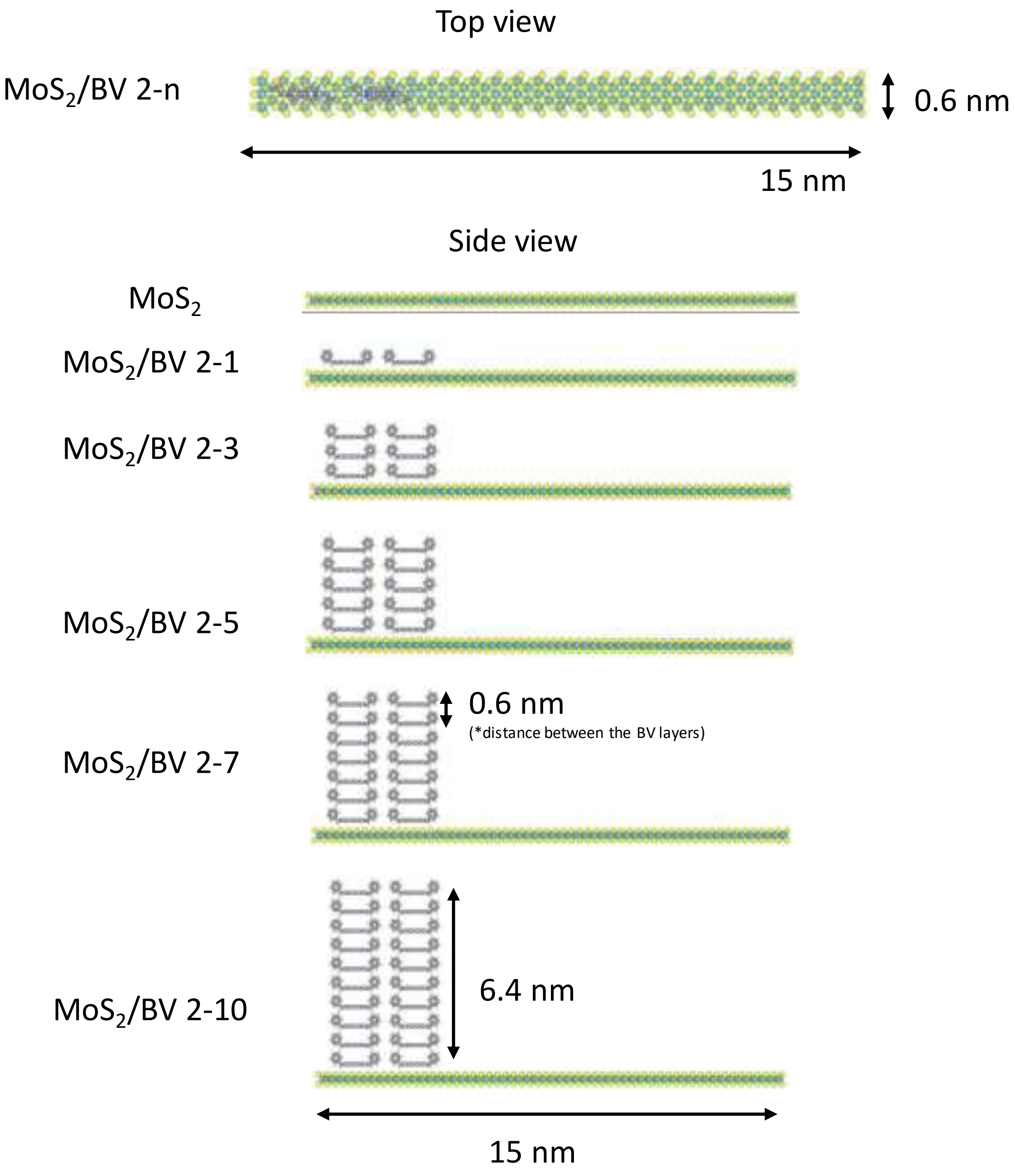

**Fig. S8**. Schematic illustration of model systems used for DFT calculations. The number of BV layers on the $MoS_2$ channel varies from 1 to 10.

Fig. S8 shows schematic illustrations of the model systems used for calculating the DOS of pristine and diffusion-doped $MoS_2$. As depicted, we modulated the number of BV molecular layers on the $MoS_2$ film. BV molecules were located only on one side of the $MoS_2$, as we applied periodic boundary conditions for the calculation.

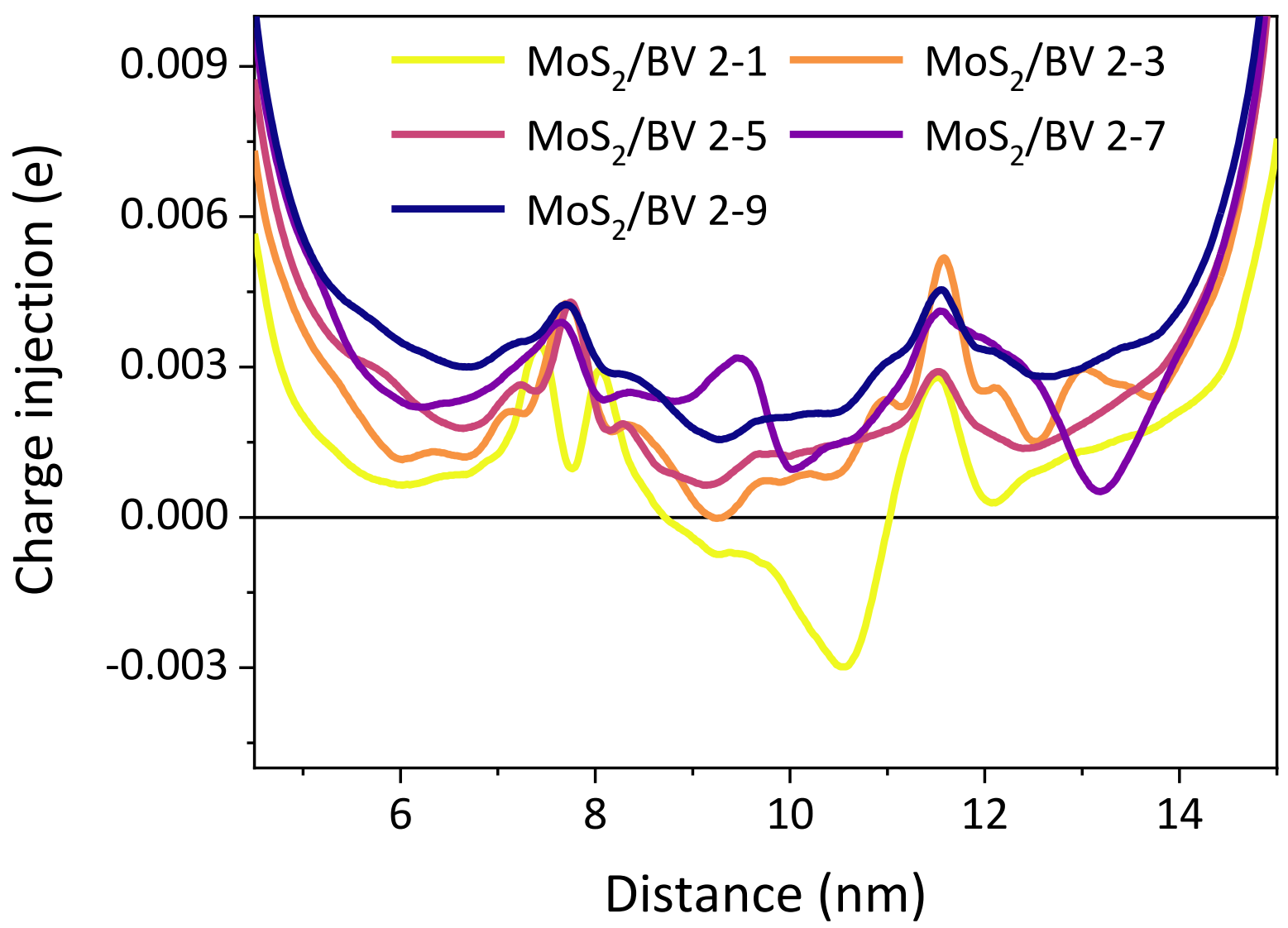

**Fig. S9**. Charge injection profile along the channel area with varying numbers of BV molecular layers.

Given such systems, we also calculated the amount of injected charges along the channel area. As shown in Fig. S9, the charge injection profile exhibited exponential decay behaviour as the distance increased from the contact area, which is consistent with experimental results. Notably, we observed that the total amount of injected charges significantly increased as the number of BV molecular layers increased.

### 3-3. $WSe_2/F_4$-TCNQ

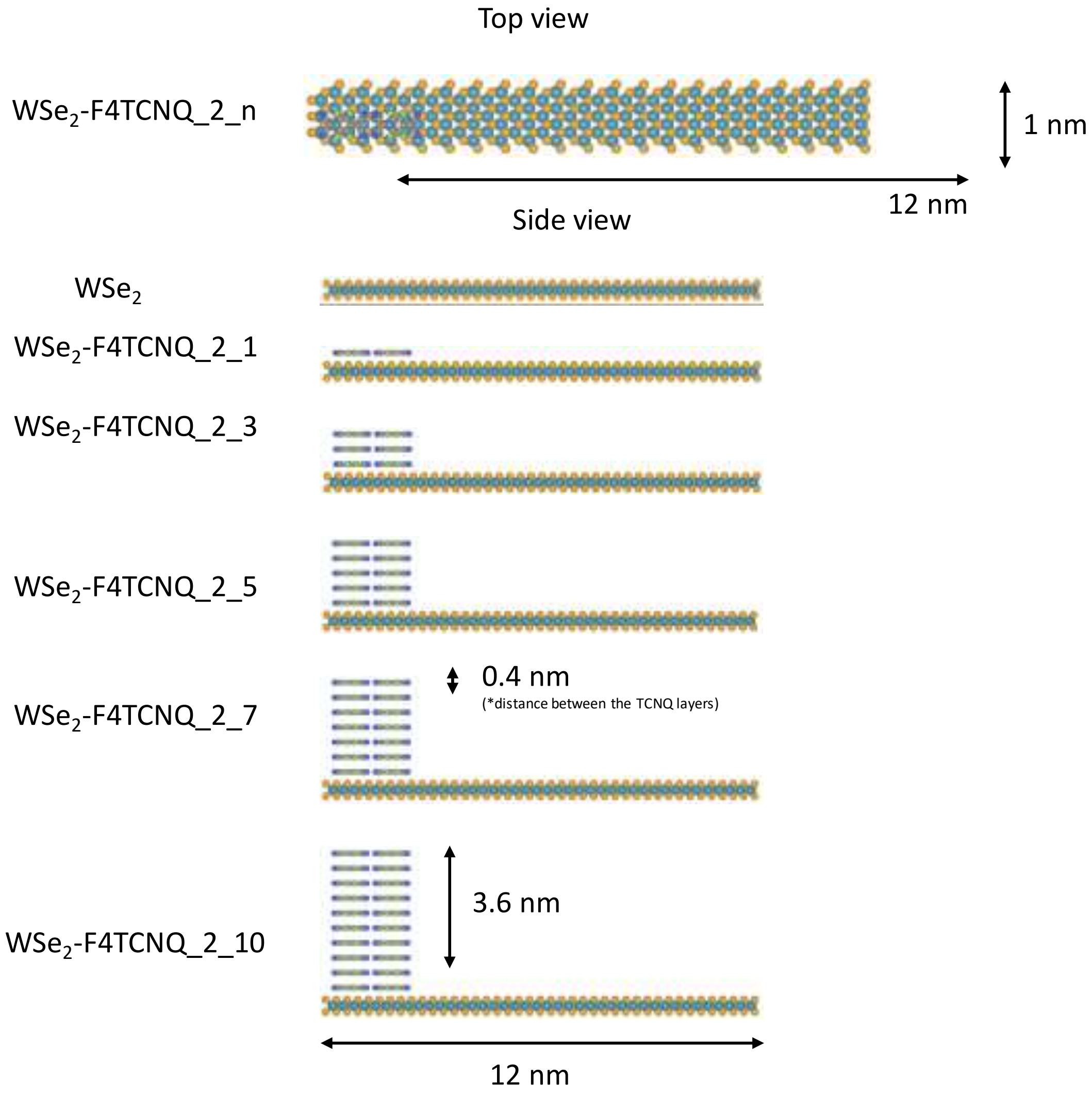

**Fig. S10**. Schematic illustration of model systems used for DFT calculations. The number of $F_4$-TCNQ layers on the $WSe_2$ channel varies from 1 to 10.

To confirm that diffusion doping method can be utilized to universal 2D layered semiconductors and molecular dopants, we also performed calculations for the tungsten diselenide ($WSe_2$) and F4-TCNQ system. $WSe_2$ exhibits ambipolar characteristics, and F4-TCNQ (2,3,5,6-Tetrafluoro-tetracyanoquinodimethane) is a representative p-type dopant widely used for SCTD. Fig. S10 shows schematic illustrations of the model systems used for these calculations.

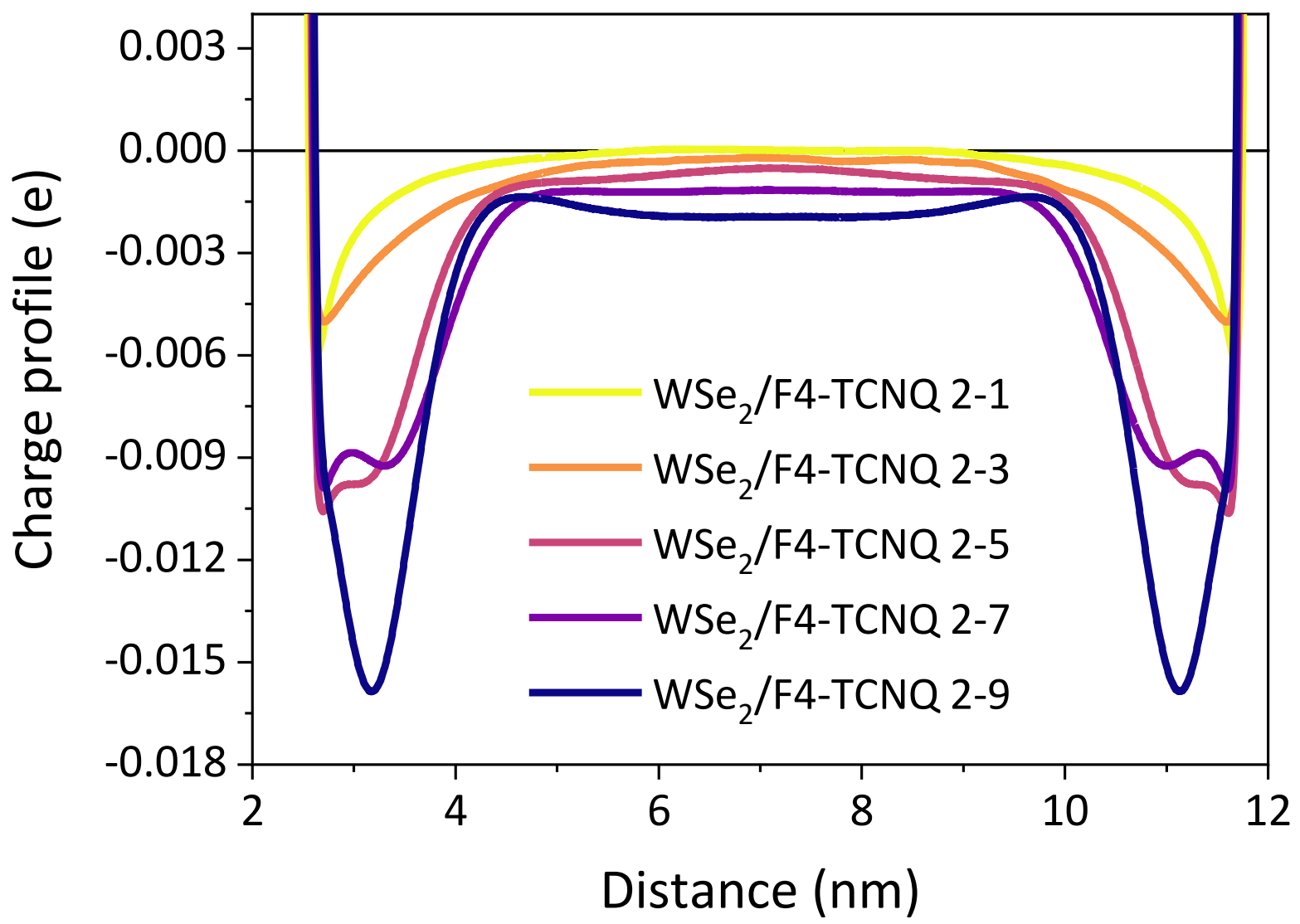

**Fig. S11**. Charge injection profile along the channel area with varying numbers of $F_4$-TCNQ molecular layers.

Since F4-TCNQ is a p-type dopant, we successfully observed hole injection throughout the channel area as shown in Fig. S11. Similarly, the total amount of injected holes increased as the number of molecular dopant layers increased. These results indicate that diffusion doping is generally effective for 2D semiconductors and molecular dopants.

## Section 4: Doping effect of CVD-$MoS_2$ FET by an immersion method

Fig. S12 displays the $V_{GS}$-$I_{DS}$ curves of a CVD-$MoS_2$ FET before (black) and after (red) BV doping, which was accomplished by immersing the device in a 5 mg/ml BV dopant solution. In this case, BV dopants are distributed throughout the entire channel area, in contrast to the diffusion doping case where BV molecules are localized only at the contact region. In particular, by comparing the off-current level and the change in carrier concentration in both cases, the effectiveness of diffusion doping could be confirmed.

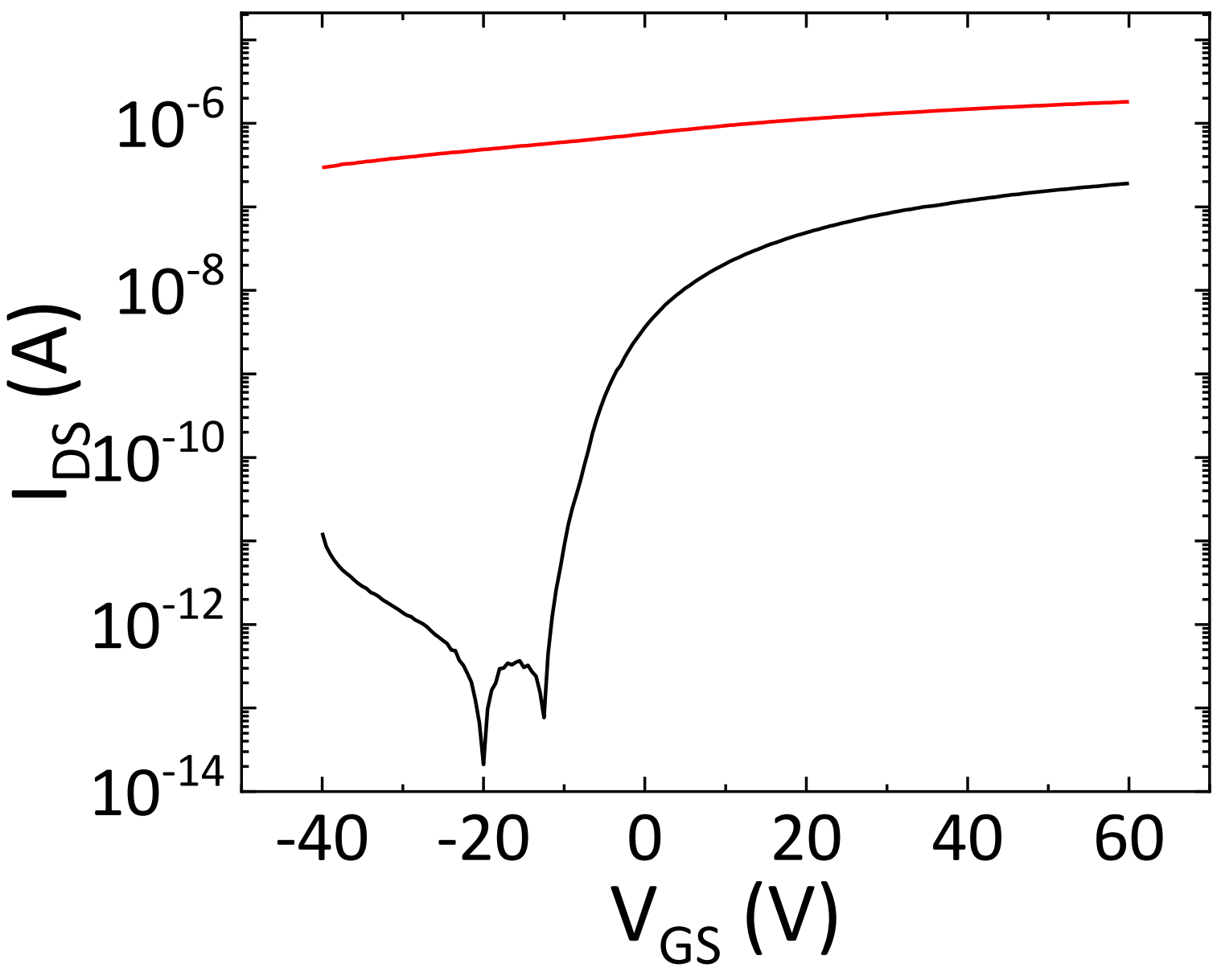

**Fig. S12**. Transfer curves of a $MoS_2$ FET before (black) and after (red) immersion doping in a 5 mg/mL BV dopant solution.

## Section 5: Low-temperature electrical characterizations

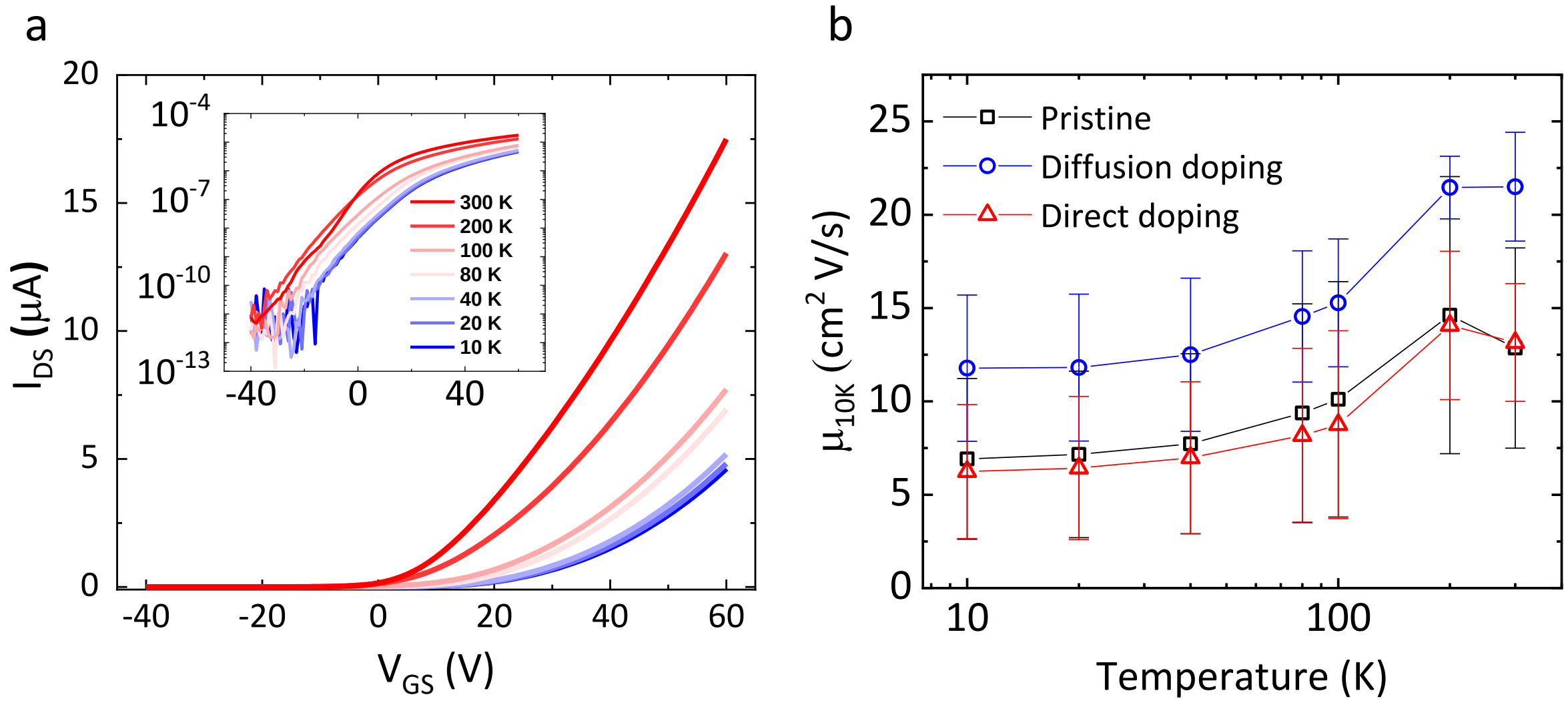

**Fig. S13**. **a**, Transfer curves of a diffusion-doped $MoS_2$ FET with varying temperatures from 300 K to 10 K. **b**, Field-effect mobilities of pristine, directly doped, and diffusion-doped $MoS_2$ FETs as a function of temperature.

To investigate the effects of charged impurities introduced by surface charge transfer dopants, namely BV molecules, we conducted temperature-dependent I-V measurements with a cryostat system (CS204*I-FMX-12, Advanced Research Systems) on pristine, directly-doped, and diffusion doped $MoS_2$ FETs. This approach is relevant because, under low-temperature conditions where phonon scattering is reduced, charged-impurity scattering becomes the key factor influencing charge mobility. As shown in Fig. S13b, while directly doped $MoS_2$ FETs displayed higher average charge mobility than pristine devices at 300 K, they exhibited lower average charge mobility at 10 K. This indicates that the increased presence of charged impurities due to BV molecules leads to reduced mobility. Conversely, the diffusion doped devices demonstrated the highest charge mobility at 10 K, suggesting less charged-impurity scattering.

## Section 6: Theoretical model for mobility calculation

We calculated charge carrier mobility at 300 K and 10 K for different charged impurity concentrations, taking into account their screening potential and scattering rate as a function of temperature and carrier concentration. Ong et al. proposed a theoretical model to calculate the charged impurity-limited mobility in TMDCs encapsulated with different dielectric materials on the top and bottom[S6]. In this model, the TMDC layer is considered as a two-dimensional electron gas (2DEG) with zero thickness. The scattering potential, induced by the single charged impurity in the $MoS_2$ channel, $\phi_q^{\mathrm{scr}}$ is given by $\phi_q^{\mathrm{scr}} = \frac{e^2 G_q}{\epsilon_{2D}(q,T)}$, where $q$ is the wave vector, $e$ is the elementary charge, and $G_q$ is the Fourier transform of the Green's function solution of the Poisson equation. The $G_q$ is expressed as $G_q = [(\epsilon_{top} + \epsilon_{bottom} \coth(q t_{ox})] \, q]^{-1}$, where $\epsilon_{top}$ is the static permittivity of the top dielectric, $\epsilon_{bottom}$ is the static permittivity of the bottom dielectric and $t_{ox}$ is the thickness of the bottom dielectric. The $G_q$ also includes the electrostatic boundary conditions.

In the direct doping case, the top dielectric is air ($\epsilon_{top} = 1$) and the bottom dielectric is 270 nm-thick $SiO_2$ ($\epsilon_{bottom} = 3.9$ and $t_{ox}$ = 270 nm). $\epsilon_{2D}(q,T)$ is the generalized static dielectric function expressed by $\epsilon_{2D}(q,T) = 1 - e^2 G_q \Pi(q,T,E_F)$, where $\Pi(q, T, E_F)$ is the static charge polarizability and $E_F$ is the chemical potential. The static charge polarizability $\Pi(q, T, E_F)$ is expressed by $\Pi(q,T,E_F) = \int_0^\infty d\mu \frac{\Pi(q,0,\mu)}{4k_B T \cosh^2(\frac{E_F-\mu}{2k_B T})}$ , where $\Pi(q,0,\mu) = \Pi(0,0,\mu)[1 - \Theta(q - 2k_F)\left\{1 - \left(\frac{2k_F}{q}\right)^2\right\}^{\frac{1}{2}}]$ with $k_F = \frac{\sqrt{2m_{eff}\mu}}{\hbar}$ and $\Pi(0,0,\mu) = -\frac{g m_{eff}}{(2\pi\hbar^2)}$. Here, we assumed an effective mass of electron 0.55 $m_e$ for $MoS_2$[S7].

The scattering rate for the single charged impurity in the $MoS_2$ channel ($\Gamma_{imp}$) is given by

$$\Gamma_{\text{imp}}(E_k) = \frac{1}{2\pi\hbar}\int d\mathbf{k}' \left|\phi^{\text{scr}}_{|\boldsymbol{k}-\boldsymbol{k}'|}\right|^2 \times (1 - cos\theta_{\mathbf{kk}'})\delta(E_{\mathbf{k}} - E_{\mathbf{k}'}), \qquad \text{(S1)}$$

where $\theta_{\mathbf{kk}'}$ is the scattering angle between the **k** and **k'** states. From $\Gamma_{\text{imp}}$, the charged impurity-limited mobility ($\mu_{\text{imp}}$) is given by

$$\mu_{\text{imp}} = \frac{e}{\pi n \hbar^2 k_B \text{T}}\int_0^{\infty} f(E)[1 - f(E)]\,(n_{imp}\Gamma_{\text{imp}}(E))^{-1} E dE, \qquad \text{(S2)}$$

where $f(E)$ is the Fermi-Dirac function, $k_B$ is the Boltzmann constant, and n is the carrier density, and $n_{imp}$ is the charged impurity density. By using this equation, the $\mu_{imp}$ of directly doped $MoS_2$ was calculated. Charged impurity concentration values were estimated by comparing them with experimental results. As shown in Fig. S14, carrier mobility decreases as the charged impurity concentration increases, both at 300 K and 10 K.

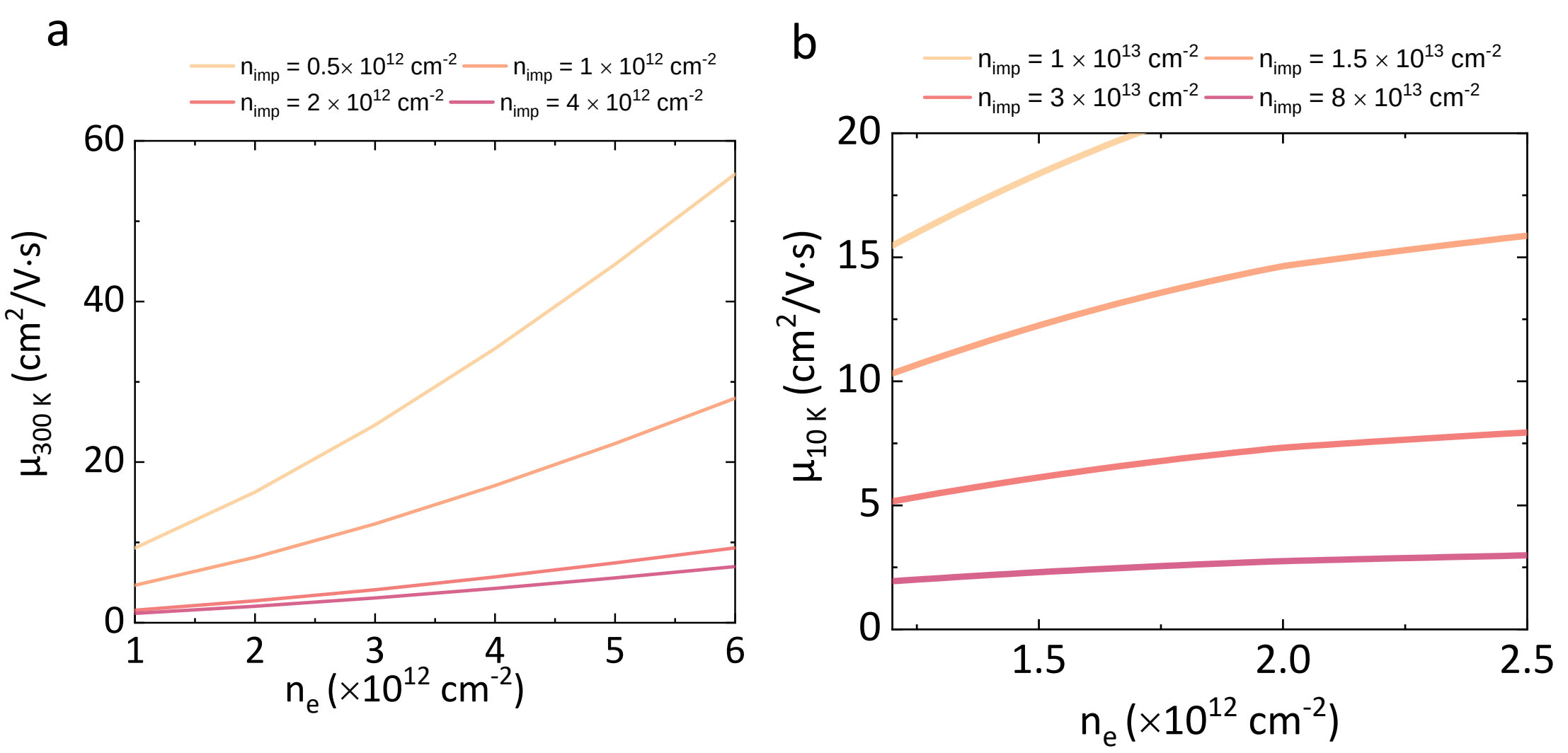

**Fig. S14.** Theoretical models for calculating the mobility of directly doped $MoS_2$ with at **a**, 300 K and **b**, 10 K with different charged impurity concentrations. Carrier mobility decreases as the charged impurity concentration increases.

**References**

S1. Perdew, J. P., Burke, K. & Ernzerhof, M. Generalized gradient approximation made simple. *Phys. Rev. Lett*. **77**, 3865-3868 (1966).

S2. Kresse, G. & Furthmuller, J. Efficient iterative schemes for ab initio total-energy calculations using a plane-wave basis set. *Phys. Rev. B* **54**, 11169-11186 (1996).

S3. Grimme, S., Ehrlich, S. & Goerigk, L. Effect of the damping function in dispersion corrected density functional theory. *J. Comput. Chem*. **32**, 1456 (2011).

S4. Becke, A. D. Density-functional thermochemistry. III. The role of exact exchange. *J. Chem. Phys*. **98**, 5648–5652 (1993).

S5. Neese, F. Software update: The ORCA program system—Version 5.0. *Wiley Interdiscip. Rev. Comput. Mol. Sci.* **12**, e1606 (2022).

S6. Ong, Z.-Y. & Fischetti, M. V. Mobility enhancement and temperature dependence in top-gated single-layer $MoS_2$. *Phys. Rev. B* **88**, 165316 (2013).

S7. Desai, S. B. *et al*, $MoS_2$ transistors with 1-nanometer gate lengths. *Science* **354,** 99–102 (2016).